\documentclass[pre,superscriptaddress,showpacs,aps]{revtex4-1}
\setcitestyle{authoryear,round}
\usepackage{graphicx}
\usepackage{dcolumn}
\usepackage{bm}
\usepackage{amsmath}
\usepackage{amssymb}
\usepackage{color}
\usepackage{epsfig}

\usepackage[T2A]{fontenc}
\usepackage[cp1251]{inputenc}

\usepackage{natbib}

\begin{document}

\bibliographystyle{apa}  

\title{Fluid theory of coherent magnetic vortices in high-$\bm\beta$ space plasmas}
\vspace{.25cm}

\author{Du\v san Jovanovi\'c}
\email{dusan.jovanovic@ipb.ac.rs}
\affiliation{Institute of Physics, University of Belgrade, Pregrevica 118, 11080 Belgrade (Zemun), Serbia}
\affiliation{State University of Novi Pazar, Vuka Karad\v{z}i\'ca bb, 36300 Novi Pazar, Serbia}
\author{Olga Alexandrova}
\email{olga.alexandrova@obspm.fr} \affiliation{LESIA, Observatoire de Paris, Universit\'{e} PSL, CNRS, Sorbonne Universit\'{e}, Universit\'{e} de Paris, 5
place Jules Janssen, 92195 Meudon, France}
\author{Milan Maksimovi\'c}
\email{milan.maksimovic@obspm.fr} \affiliation{LESIA, Observatoire de Paris, Universit\'{e} PSL, CNRS, Sorbonne Universit\'{e}, Universit\'{e} de Paris, 5
place Jules Janssen, 92195 Meudon, France}
\author{Milivoj Beli\'c}
\email{milivoj.belic@qatar.tamu.edu} \affiliation{Texas A\&M University at Qatar, P.O. Box 23874 Doha, Qatar}
\date{\today}
\begin{abstract}

In-situ observations in the Earth's and Saturn's magnetosheaths and in the solar wind reveal the presence of Alfv\'en vortices as intermittent structures in the range of scales from fluid lengths down to few ion lengths. The density and the magnetic field associated with them appear to be compressible for higher plasma betas. Until now, only incompressible Alfv\'en vortices have been known. Motivated by space plasma observations we develop a new model of magnetic vortices in high-beta plasmas with anisotropic temperature, possessing compressible density and magnetic field, whose typical size ranges from fluid to ion scales. At magneto-fluid scales we find novel non-propagating field-aligned cylindrical monopoles and inclined propagating dipoles. Their transverse magnetic and velocity fluctuations are aligned, but not identical, {and they exhibit density and compressible magnetic field fluctuations $\delta n$ and  $\delta B_\Vert$ localized inside the vortex core. In the presence of thermal anisotropy and acoustic effects, they may be correlated or anti-correlated $\delta n/\delta B_\Vert={\rm constant}\gtrless 0$; fluctuations whose velocity along the magnetic field is below the ion thermal speed  are always correlated.} At ion or kinetic scales (with the smallest radii $\sim c/\omega_{pi}, \rho_{L i}$) {and in the absence of acoustic perturbations}, only dipolar Alfv\'en vortices survive with properties similar as those at fluid scales, except that $\delta n/n_0$ reaches the level of $\delta B_\Vert/B_0$.
{We also find pressure balanced kinetic slow magnetosonic dipoles, possessing finite 
$E_\Vert$ and purely compressional magnetic field perturbation, whose existence is facilitated by a strong ion temperature anisotropy.}

\end{abstract}

\pacs{
52.35.Ra, 
52.35.We, 
94.05.Lk, 
94.30.cj, 
96.50.Ci  
}
\maketitle

\section{Introduction}\label{Introductory}

Magnetic structures at ion break scale, commonly in the form of Alfv\'en vortices with the diameter $10-30$ times longer than the ion scales, have been observed in the solar wind and in the magnetosheaths of the Earth and Saturn \citep{2004JGRA..109.5207A,2008GeoRL..3515102A,2006JGRA..11112208A,2008NPGeo..15...95A,Denises_paper,2017ApJ...849...49P,2016ApJ...824...47L}. {Similar structures, but with the diameter comparable to the ion-acoustic Larmor radius and identified as the drift-Alfv\'{e}n vortices, were observed in the Earth's magnetospheric cusp region \citep{2005Natur.436..825S} in which the ratio of thermal and magnetic pressures is considerably smaller.} A detailed statistical analysis of diverse magnetic fluctuations in the turbulent cascade close to the ion spectral break, detected in the slow and fast solar wind streams by the multi-spacecraft Cluster mission has been presented by
\citet{Denises_paper,2017ApJ...849...49P}. They have shown that the intermittency of the magnetic turbulence is due to the presence of coherent structures of various nature. The compressible structures observed in the slow wind are predominantly parallel perturbations of the magnetic field ($\delta B_\Vert\gg\delta B_\bot$) and have the form of magnetic holes, solitons and shock waves. Coherent shear Alfv\'enic perturbations have been observed both in the slow and the fast solar wind, featuring $\beta_i\gtrsim 1$ and $\beta_i\lesssim 1$, respectively, where $\beta = 2 p/c^2 \epsilon_0 B^2$ is the ratio between the plasma pressure $p$ and the magnetic pressure. They appear either as the current sheets or the vortex-like structures. Predominantly torsional Alfv\'enic vortices with $\delta B_\bot \gg\delta B_\Vert$, but with finite $\delta B_\Vert$, are commonly present both in the slow and the fast wind. Vortices with a larger compressional magnetic field component, $\delta B_\bot \gtrsim\delta B_\Vert$, have been observed only in the slow wind. The observed compressible component $\delta B_\Vert$ is usually well localized within the structure, while the torsional part $\delta B_\bot$ is more delocalized, extending itself outside of the vortex core.

The multi-point Cluster measurements have enabled the determination of the spatial and temporal properties of these structures, such as their propagation velocity, the direction of the normal, and the spatial scale along this normal. The normal is always perpendicular to the local magnetic field, indicating that the structures are strongly elongated in the direction of $\vec B$. The majority of the structures is convected by the wind, but in the slow wind a significant fraction ($\sim 25\%$) propagates in the perpendicular direction and with finite velocities relative to the plasma. In the fast wind, no structures propagating relative to the plasma could be verified because the propagation velocities that came out were smaller than the error of measurements. Typical scales of the structures along the normal are $2-5$ characteristic ion lengths, such as the ion Larmor radius $v_{T i_\bot}/\Omega_i$, acoustic radius $c_S/\Omega_i$, and ion inertial length $c/\omega_{pi}$, that in a solar wind plasma are of the same order of magnitude.
%
%
%
%
%
%
Similar features of plasma turbulence were observed previously in the magnetosheaths 
downstream of quasi-perpendicular bow shocks of the Earth and Saturn by 
\citep{2004JGRA..109.5207A,2006JGRA..11112208A,2008GeoRL..3515102A}, who detected coherent shear Alfv\'enic vortex structures in the form of current filaments slightly tilted relative to the 
magnetic field, $\nabla_\Vert\ll \nabla_\bot$, exhibiting only perpendicular magnetic perturbations, $\delta B_\Vert\to 0$.
Recently, using the high resolution in-situ measurements from the Magnetospheric Multiscale (MMS) mission, \citet{2019ApJ...871L..22W} presented a detailed analysis of the plasma features within an Alfv\'{e}n vortex. They demonstrated that the quasi-monopolar, mostly torsional $\nabla_\Vert\ll \nabla_\bot$, Alfv\'{e}n   vortex with a radius of $\sim\!\! 10$ proton Larmor radii observed in the Earth's turbulent magnetosheath had the magnetic fluctuations $\delta \vec{B}_\bot$ anti-correlated from the velocity fluctuations $\delta \vec{V}_\bot$, while its compressive features were in a qualitative agreement with the theory developed in the present paper.

Shear Alfv\'enic fluid vortices were predicted theoretically by 
\citet{1992swpa.book.....P}, who demonstrated that structures of this type with the transverse size bigger than the ion inertial length $c/\omega_{pi}$, where $\omega_{pi} = (n_0 e^2/ m_i \epsilon_0)^{1/2}$ is the ion plasma frequency, could exist in plasmas with incompressible density (usually occurring when $\beta$ is small). Under such conditions, nonlinear solutions are described by the standard Kadomtsev-Pogutse-Strauss' reduced MHD (magnetohydrodynamic) equations \citep{1973ZhETF..65..575K,1974ZhETF...66.2056K,
1976PhFl...19..134S,1977PhFl...20.1354S}.
Solutions can be nontravelling monopoles or propagating structures. A hydrodynamic vortex that moves relative to the plasma always has the form of a dipole, but also a monopole can be superimposed to it. We will show below that such monopolar component is possible only in the absence of the compressibility of the magnetic field, that is realized if either the structures are (much) bigger than the ion scale, or the plasma $\beta$ is sufficiently small.
O. Alexandrova proposed \citep{2006JGRA..11112208A,2008NPGeo..15...95A} that, within the Kadomtsev-Pogutse-Strauss' reduced MHD description, such vortices might be created 
by the filamentation of the nonlinear slab-like structures arising from the saturation of the linearly unstable Alfv\'{e}n ion cyclotron waves \citep{2004JGRA..109.5207A} or, more likely, arising naturally as the intermittency of the turbulence \citep{2006JGRA..11112208A,2008NPGeo..15...95A,Denises_paper,2017ApJ...849...49P,2016ApJ...824...47L,2016JGRA..121.3870R}.
When a spacecraft encounters a dipole, the recorded signal depends on the relative position of the dipole's axis and the satellite's trajectory. The satellite may observe either one "pole" of the dipole or both, and the detected signals superficially appear to be qualitatively different.


In this paper we present a hydrodynamic theory of coherent vortices in the space plasmas 
that can be characterized with anisotropic electron and ion temperatures, and with arbitrary plasma $\beta$.
We generalize the classical shear-Alfv\'{e}n result 
\citep{1992swpa.book.....P}
by including the diamagnetic and finite Larmor radius effects via Braginskii's collisionless stress tensor, and  the compressional magnetic component via a generalized pressure balance. We demonstrate that perturbations that are bigger that the ion inertial length are properly described by the Kadomtsev-Pogutse-Strauss' equations of reduced MHD and that in plasmas with a modest $\beta\sim 1$, such description remains valid also 
when the size of the structure is only slightly bigger than the ion inertial length. We find a general reduced MHD vortex solution, whose torsional component of the magnetic field is leaking outside of the vortex core while the compressional magnetic field is restricted to its interior, which is why the latter may remain undetected by a spacecraft. They possess also a finite density perturbation and parallel fluid velocity that are proportional to the vorticity and the compressional magnetic field. Furthermore, in plasmas with $\beta\lesssim 1$ and on a characteristic length that belongs to the ion-scale, we find two different particular solutions in the form of dipole vortices that can be regarded as the generalizations of the Kadomtsev-Pogutse-Strauss' structures to the smaller scales, and of the nonlinear drift-mode to the slow mode structures in large-$\beta$ plasmas, respectively. Our generalized Kadomtsev-Pogutse-Strauss' dipoles possess all three components of the magnetic field perturbation and their phase velocity component in the direction of the ambient magnetic field lies in the Alfv\'en and the acoustic ranges, $u_\Vert\sim (c_A, c_S)$, {while the nonpropagating monopoles have $u_\Vert\to\infty$}. Thee slow magnetosonic dipoles that we are able to study analytically propagate much slower, $u_\Vert\ll c_A$ (the range of permitted $u_\Vert$ is broadened in the presence of ion temperature anisotropy) and their magnetic field perturbation is mostly compressional, i.e. $\vec{B}_\bot\to 0$. The 
moving monopolar structures in the compressional magnetic field will be considered in a separate publication, since they require a (gyro)kinetic description due to their ability to trap particles in the parallel direction and to redistribute their parallel and perpendicular temperatures.

\section{Fluid theory of slow, weakly $\bm z$-dependent nonlinear phenomena in a warm plasma $\bm{\beta_{i_\bot} \sim \beta_{e_\bot}\sim 1}$}\label{hydro_derivation}

We consider a collisionless plasma with the unperturbed density $n_0$ immersed in the homogeneous magnetic field $\vec{e}_z B_0$. We assume that the electron and ion temperatures can be anisotropic, i.e. that the temperatures associated with the particles' random motions along and perpendicularly to the magnetic field may be different, $T_{j_\Vert} \ne T_{j_\bot}$, where $j =  e, i$ and the subscripts $\Vert$ and $\bot$ denote the components parallel and perpendicular to the magnetic field, respectively. The hydrodynamic equations of continuity and momentum 
for each of the species have the form
\begin{eqnarray}\label{continuity_eq}
&& \left(\frac{\partial}{\partial t} + \vec{U}\cdot\nabla\right)n + n\nabla\cdot\vec{U} = 0, \\
\label{momentum_eq}
&&  \left(\frac{\partial}{\partial t} + \vec{U}\cdot\nabla\right)\vec{U} = \frac{q}{m}\left(\vec{E} + \vec{U}\times\vec{B}\right) - \frac{1}{m n}\nabla\cdot\left(\bm{P} + \bm{\pi}\right),
\end{eqnarray}
where, for simplicity, we have omitted the subscripts $e$ and $i$ referring to the electrons and ions, respectively. In the above, $n$, $\vec{U}$, $q$, and $m$ are the number density, fluid velocity, charge, and mass, respectively. The pressure $\bm P$ and the stress $\bm \pi$ are diagonal and off-diagonal tensors. Using the standard shorthand notation, the pressure tensor in the case of an anisotropic temperature is given by $\bm{P} = p_\bot(\bm{I} - \vec{b}\,\vec{b}) + p_\Vert\vec{b}\,\vec{b}$ where $\bm{I}$ is a unit tensor, viz. $I_{\alpha, \beta} = \delta_{\alpha, \beta}$ and $\delta_{\alpha, \beta}$ is the Kronecker delta, {and $\vec{b}\,\vec{b}$ stands for the dyadic product, whose components are given by $(\vec{b}\,\vec{b})_{\alpha,\beta}=b_\alpha b_\beta$. Here} $\vec{b}$ is the unit vector parallel to the magnetic field, $\vec{b} = \vec{B}/B$, $B$ is the intensity of the magnetic field, $B = |\vec{B}|$, $p_\Vert = n T_\Vert$, and $p_\bot = n T_\bot$. Likewise, the stress tensor is written as  $\bm{\pi} = \vec{e}_\alpha \vec{e}_\beta \; \pi_{\alpha,\beta}$. These enable us to use the standard formula from the tensor algebra 
$\nabla\cdot\vec{q} \, \vec{r} = (\nabla\cdot\vec{q} + \vec{q}\cdot\nabla)\vec{r}$, and to write the divergence of the pressure and the stress tensors as
\begin{eqnarray}
\label{pressure_divergence}
&&
\nabla\cdot\bm{P}={\nabla p_\bot +\vec{b}\left(\vec{b}\cdot\nabla\right)\left(p_\Vert-p_\bot\right)+ \left(p_\Vert-p_\bot\right)\left(\nabla\cdot\vec{b} + \vec{b}\cdot\nabla\right)\vec{b}},
\\
\label{tensor_divergence}
&&
\nabla\cdot\bm{\pi}
= \vec{e}_\beta\left(\vec{e}_\alpha\cdot\nabla\right) \pi_{\alpha, \beta}
+ \pi_{\alpha,\beta}\left(\nabla\cdot\vec{e}_\alpha + \vec{e}_\alpha\cdot\nabla\right)\vec{e}_\beta.
\end{eqnarray}
\textcolor{black}{
The endmost terms on the right-hand-sides on the above equations arise from the curvature of magnetic field lines. For later convenience, we introduce the notation
\begin{equation}\label{curvature}
\left(\nabla\cdot\bm{\pi}\right)_{curv} = \pi_{\alpha,\beta}\left(\nabla\cdot\vec{e}_\alpha + \vec{e}_\alpha\cdot\nabla\right)\vec{e}_\beta.
\end{equation}}
The chain of hydrodynamic equations is truncated by using the \citet{1965RvPP....1..205B} collisionless (nongyrotropic) stress tensor, appropriate for perturbations 
weakly varying both on the timescale of the gyroperiod and on the perpendicular scale of the Larmor radius. Within the adopted large scale limit, Braginskii's stress tensor has been generalized to anizotropic temperatures, see e.g. classical works by \citet{1966PThPh..36....1Y} and the more recent ones by \citet{2010MNRAS.405..291S} and \citet{2012AIPC.1439...94S,2015JPlPh..81a3203S}. Following these authors who neglected the heat flux, we disregard the hydrodynamic equations for pressure and stress tensors, and use instead the generic equations of state
\begin{equation}\label{pressure}
d p_\bot/p_\bot = \gamma_\bot \; dn/n , \quad\quad d p_\Vert/p_\Vert = \gamma_\Vert \; dn/n ,
\end{equation}
in which the multipliers $\gamma_\Vert$ and $\gamma_\bot$ are some functionals of the plasma density, see also Refs. \citep{1992JGR....97.8327B,passot1} where the original (complex) polytropic indices for collisionless plasmas have been derived. We consider regimes in which the perturbations of the density and of the magnetic field are not too large, see Eq. (\ref{weak_z_dep}), which permits us to make an estimate of the functionals $\gamma_\Vert$ and $\gamma_\bot$ from the linearized Vlasov equation. These are further simplified under the drift scaling (\ref{drift_scaling}) and for weak dependence along magnetic field lines, (\ref{weak_z_dep}). Under such conditions, the parallel functional $\gamma_\Vert$ reduces to a constant, viz. $\gamma_\Vert = 3$ when the characteristic parallel velocity of propagation (i.e. the phase velocity $u_z$) is bigger than the parallel thermal velocity $v_{T_\Vert}$ and the process can be considered as adiabatic, and to $\gamma_\Vert = 1$ when $u_z\ll v_{T_\Vert}$, i.e. the process is isothermal.
Likewise, the perpendicular functional $\gamma_\bot$ reduces to $\gamma_\bot = 1$  for arbitrary ratios $u_z/v_{T_\Vert}$ if the characteristic perpendicular size of the solution is much bigger than the Larmor radius. Conversely, for solutions whose transverse scale approaches the ion scales, $\gamma_\bot$ can be approximated by a constant only in a limited number of cases, for which vortex solutions are found in Section \ref{Traveling}. These are the 
shear Alfv\'{e}n solution whose parallel electric field is zero $E_\Vert = 0$,
decoupled from acoustic perturbations, $u_z> v_{T i_{\Vert,\bot}}$, and the 
kinetic slow magnetosonic mode solution whose torsion of the magnetic field is absent, $\delta\vec{B}_\bot=0$, realized in the regime $c_A> u_z> v_{T i_{\Vert,\bot}}$, when the coupling with acoustic perturbation is negligible. In both cases we have $\gamma_{i_\bot} = 2 $.
For more details, see Appendix \ref{eqstate}.

The system of equations is closed by the Faraday's and Ampere's laws
\begin{equation}
\label{Maxwell} \nabla\times\vec{E} = -\frac{\partial \vec{B}}{\partial t}, \quad \quad
\nabla\times\vec{B} = \frac{1}{c^2}\left(\frac{\partial \vec{E}}{\partial t} + \frac{\vec{j}}{\epsilon_0}\right).
\end{equation}
The latter is simplified on temporal scales that are slow compared to the electron plasma frequency $\omega_{p e} = \sqrt{n_0 e^2/m_e\epsilon_0}$ and spatial scales
that are long compared to the electron Debye length $\lambda_{D e} = v_{T e}/\omega_{p e}$, when we can neglect both the charge separation and the displacement current, yielding
\begin{equation}\label{quasineutr_no_displac}
n_e = n_i \equiv n, \quad\quad {\frac{n}{n_0}}\left(\vec{U}_i - \vec{U}_e\right) = \frac{e}{m_e}\frac{c^2}{\omega_{p e}^2} \; \nabla\times\vec{B}.
\end{equation}
Here and in the rest of the paper, we consider single-charged ions, viz. $q_i = -q_e = e$.

For later reference, we write also the parallel momentum equation, that is obtained when we multiply the momentum equation with $\vec{b} \; \cdot$ , viz.
\begin{equation}\label{parallel_momentum_eq}
\left(\frac{\partial}{\partial t} + \vec{U}\cdot\nabla\right)U_\Vert -
\frac{q}{m} \; \vec{b}\cdot\left(\vec{E} - \frac{1}{qn} \; \nabla p_\Vert + \frac{p_\Vert - p_\bot}{qnB} \; \nabla B\right) +
\frac{1}{m n}\left[ \frac{\partial\pi_{\alpha,b}}{\partial x_\alpha}+ \vec{b}\cdot \left(\nabla\cdot\bm{\pi}\right)_{curv}\right] =
\vec{U}_\bot\cdot\left(\frac{\partial}{\partial t} + \vec{U}\cdot\nabla\right)\vec{b},
\end{equation}
where $U_\Vert = \vec{b}\cdot\vec{U}$. Likewise, multiplying the electron- and ion momentum equations by $m_e n_e$ and $m_i n_i$, respectively, adding, and taking the component perpendicular to the magnetic field, after some tedious but straightforward algebra, we obtain the perpendicular momentum equation for the plasma fluid
\begin{eqnarray}\label{MHD_momentum}
\nonumber
&& \nabla_\bot\left(c^2 \epsilon_0 \frac{B^2}{2} + p_{e_\bot} + p_{i_\bot}\right) = \vec{b}\times\left\{\vec{b}\times\left[m_e n_e\left(\frac{\partial}{\partial t} + \vec{U}_e\cdot\nabla\right)\vec{U}_e + m_i n_i\left(\frac{\partial}{\partial t} + \vec{U}_i\cdot\nabla\right)\vec{U}_i\right]\right\} + \\
&&
\epsilon_0\left[\vec{E}_\bot\left(\nabla\cdot \vec{E}\right) + B\vec{b}\times\frac{\partial \vec{E}}{\partial t}\right] + \left(c^2 \epsilon_0 B^2 - p_{e_\Vert} - p_{i_\Vert} + p_{e_\bot} + p_{i_\bot}\right)\left(\vec{b}\cdot\nabla\right)\vec{b} -
\nabla\cdot\left(\pi_{e_{\alpha,\beta}} + \pi_{i_{\alpha,\beta}}\right),
\end{eqnarray}
{that is in the quasineutrality regime, Eq.(\ref{quasineutr_no_displac}) {(i.e. on the adopted scales bigger than the Debye length $\lambda_{De}$)}, simplified by setting $\nabla\cdot\vec{E}= e(n_i-n_e)/\epsilon_0=0$.} Our equations (\ref{continuity_eq}), (\ref{momentum_eq}), (\ref{parallel_momentum_eq}), and (\ref{MHD_momentum}) are vastly simplified under the drift scaling
\begin{equation}\label{drift_scaling}
\frac{1}{\Omega}\frac{\partial}{\partial t} \sim \frac{1}{\Omega} \, \vec{U}\cdot\nabla \sim  \epsilon \ll 1,
\end{equation}
($\epsilon$ being a small parameter) and in the regime of small perturbations of the density and of the magnetic field, and of the weak dependence along the magnetic field line
\begin{equation}\label{weak_z_dep}
\frac{\delta n}{n} \sim \frac{|\delta\vec{B}|}{|\vec{B}|} \sim \frac{\vec{b} \, \cdot\nabla}{\nabla_\bot} \sim \epsilon,
\end{equation}
provided the fluid motion is not predominantly 1-d in the parallel direction, $\vec{U}_\Vert \not\gg \vec{U}_\bot$. Here, $\Omega$ is the gyrofrequency, $\Omega= q B/m$, and $\delta$ denotes the deviation of a quantity from its unperturbed value.
The explicit form of the collisionless stress tensor $\bm\pi$ in a plasma with anisotropic temperature can be seen e.g. in Refs. \citep{1966PThPh..36....1Y,sulem1,passot1},
where it has been calculated under the scaling (\ref{drift_scaling}), with the accuracy to the the first order in the FLR (finite-Larmor-radius) corrections, and expressed in the local orthogonal coordinate system with the curvilinear b-axis along the magnetic lines of force
\begin{eqnarray}
\nonumber && \pi_{m,m} = - \pi_{l,l} = \left(p_\bot/2\Omega\right)\left(\partial U_l/\partial x_m + \partial U_m/\partial x_l\right), \\
\nonumber && \pi_{l,m} = \pi_{m,l} = \left(p_\bot/2\Omega\right)\left(\partial U_l/\partial x_l - \partial U_m/\partial x_m\right), \\
\nonumber && \pi_{l,b} = \pi_{b,l} = -\left(p_\bot/\Omega\right)\left(\partial U_b/\partial x_{m}\right) -
\left[\left(2 p_\Vert - p_\bot\right)/\Omega\right]\left(\partial U_m/\partial x_b\right), \\
\nonumber && \pi_{m,b} = \pi_{b,m} = \left(p_\bot/\Omega\right)\left(\partial U_b/\partial x_l\right) +
\left[\left(2 p_\Vert - p_\bot\right)/\Omega\right]\left(\partial U_l/\partial x_b\right) , \\
&& \pi_{b,b} = 0.
\label{braginskii}
\end{eqnarray}
Here $\vec{b}$, $\vec{e}_l$, and $\vec{e}_m$ are three mutually perpendicular unit vectors {and $\partial/\partial x_\alpha\equiv \vec{e}_\alpha\cdot\nabla$, where $\alpha=l,m,b$}. We adopt the last two vectors as $\vec{e}_l = \vec{b}\times(\vec{e}_x\times\vec{b})/|\vec{e}_x\times\vec{b}|$ and $\vec{e}_m = \vec{b}\times \vec{e}_l$. In the regime of a weak curvature of magnetic field lines, Eq. (\ref{weak_z_dep}), these unit vectors are properly approximated by the expressions given in Eq. (\ref{unit_vectors}) below.

The perpendicular component of the fluid velocity is obtained after we multiply Eq. (\ref{momentum_eq}) with $\vec{b}\times$, and can be readily written as the sum of the $\vec{E}\times\vec{B}$, diamagnetic, anisotropic-temperature, and polarization drifts, together with the leading part and the curvature correction of the stress-related (or the FLR) drift:
\begin{equation}\label{drift_velocity}
\vec{U}_\bot = \vec{U}_E + \vec{U}_D + \vec{U}_A  + \vec{U}_p + \vec{U}_\pi + \delta\vec{U}_\pi,
\end{equation}
where
\begin{eqnarray}
&&
\label{drifts_big}
\vec{U}_E = -\frac{\vec{b}}{B}\times\vec{E}, \quad
\vec{U}_D = \frac{\vec{b}}{q n B}\times\nabla_\bot p_\bot, \quad
\vec{U}_A = \left(p_\Vert - p_\bot\right)\frac{\vec{b}}{q n B}\times\left(\vec{b}\cdot\nabla\right)\vec{b},
\quad
\vec{U}_B = \frac{p_\bot \vec{b}}{q n B^2}\times\nabla_\bot B
\\
&&
\label{drifts_small}
\vec{U}_p = \frac{\vec{b}}{\Omega}\times\left(\frac{\partial}{\partial t} + \vec{U}\cdot\nabla\right)\vec{U},
\quad
\vec{U}_\pi = \frac{\vec{b}}{q n B}\times\vec{e}_{\beta}\frac{\partial\pi_{\alpha, \beta}}{\partial x_{\alpha}},
\quad
\delta\vec{U}_\pi = \frac{\vec{b}}{q n B}\times\left(\nabla\cdot\bm{\pi}\right)_{curv}.
\end{eqnarray}
For completeness, in the above list we have added also the grad-$B$ drift velocity $\vec{U}_B$, although it does not appear explicitly in the expression (\ref{drift_velocity}), but it will turn up later in the continuity equation, by virtue of the term $\nabla\cdot\vec{U}$.

Using 
expressions Eq. (\ref{braginskii}), the stress-related drift and the leading contribution of the stress to the parallel momentum equation (\ref{parallel_momentum_eq}) take the form: 
\begin{eqnarray}
\nonumber
&&\hspace{-.3cm}\vec{U}_\pi =  - \frac{T_\bot}{2 m\Omega^2}\nabla^2_\bot\vec{U}_\bot -
\\
&&\hspace{.4cm}
\frac{1}{q n B}\left\{\left[\left(\vec{b}\times\nabla_\bot\frac{p_\bot}{2\Omega}\right)\cdot\nabla_\bot\right]
\vec{b}\times\vec{U}_\bot + \left(\nabla_\bot\frac{p_\bot}{2\Omega}\cdot\nabla_\bot\right)\vec{U}_\bot +
\left(\vec{b}\cdot\nabla\right)
\left[\frac{p_\bot}{\Omega}\nabla_\bot U_\Vert +
\frac{2 p_\Vert-p_\bot}{\Omega}\, \left(\vec{b}\cdot\nabla\right)\vec{U}_\bot\right]
\right\}, \quad\quad\quad
\label{U_pi}
\\
&&\hspace{-.3cm}\frac{\partial\pi_{\alpha,b}}{\partial x_\alpha} =
\vec{b}\cdot\left\{\nabla_\bot U_\Vert\times\nabla_\bot\frac{p_\bot}{\Omega} -
\nabla_\bot\times\left[\frac{2 p_\Vert-p_\bot}{\Omega} \left(\vec{b}\cdot\nabla\right)\vec{U}_\bot\right]\right\}. \label{contr_stress}
\end{eqnarray}
In the regime of slow variations (compared to $\Omega$), weak dependence along the magnetic field lines, and small perturbations of the magnetic field in all spatial directions, that is described by the scalings Eqs. (\ref{drift_scaling}) and (\ref{weak_z_dep}), {we can neglect the last term in Eq. (\ref{U_pi}) that contains the second parallel derivative $(\partial/\partial x_b)^2$}. Noting also that the contribution of the magnetic curvature to the divergence of the stress tensor is a small quantity of the order $\epsilon^2\rho_{L}^2\nabla^2$, viz.
\begin{equation}
\left(\nabla\cdot\bm{\pi}\right)_{curv} =
\pi_{\alpha,b}\;\left({\partial\vec{b}}/{\partial x_\alpha}\right) - \vec{b}\;\pi_{\alpha,\beta}\left[\vec{e}_\beta\cdot\left({\partial\vec{b}}/{\partial x_\alpha}\right)\right]
+ {\cal O}\left(\epsilon^3\right),
\label{curvature2}
\end{equation}
and using Eqs. (\ref{braginskii}) {and the approximations in Eq. (\ref{unit_vectors})}, we obtain the following leading-order expressions:
\begin{equation}\label{curvature3}
\delta\vec{U}_\pi =
{ \vec{e}_{\alpha_\bot} \, \rho_{L}^2\left(\frac{\partial\vec{b}}{\partial x_{\alpha_\bot}}\cdot\nabla U_\Vert\right)},
\quad\quad {\rm and} \quad\quad
\vec{b}\cdot \left(\nabla\cdot\bm{\pi}\right)_{curv} = q  n_0 \rho_{L}^2 \; \frac{\partial\vec{U}_\bot}{\partial x_{\alpha_\bot}}\cdot\nabla_\bot\frac{\partial A_z}{\partial x_{\alpha_\bot}},
\end{equation}
where $\rho_{L} = \sqrt{T_\bot/m\Omega_0^2}$  is the (unperturbed) Larmor radius and $\alpha_\bot = x,y$.

The leading order expression (in $\epsilon$) for the stress-related drift $\vec{U}_\pi +\delta\vec{U}_\pi$ is given by the first term on the right-hand-side of Eq. (\ref{U_pi}). Then, using the facts that the polarization drift $\vec{U}_p$ and the curvature related drifts  $\vec{U}_A$ and $\delta\vec{U}_\pi$ are small
compared to the $\vec{E}\times\vec{B}$ and diamagnetic drifts {and noting that $\nabla\cdot(\vec{U}_E+\vec{U}_D) = {\cal O}\left(\epsilon^2\right)$, see Eq. (\ref{div_U_E_U_D})}, with accuracy to leading order in $\epsilon$ we can write
$\vec{U}_\bot \approx \vec{U}_E + \vec{U}_D - \rho_{L}^2\nabla^2_\bot\vec{U}_\bot/2$,
that is formally solved as $\vec{U}_\bot \approx \vec{U}_\bot^{apr}$, where
\begin{equation}\label{U_prblizno}
\vec{U}_\bot^{apr}  \equiv\left(\textcolor{black}{1 + \rho_{L}^2\nabla^2_\bot/2}\right)^{-1}\left(\vec{U}_E + \vec{U}_D\right).
\end{equation}
%
Within the adopted accuracy, this expression for $\vec{U}_\bot$ can be used on the right-hand-sides of Eqs. (\ref{drifts_small})--(\ref{contr_stress}) and (\ref{pres_bal}).
Likewise, in the convective derivatives we use $\vec{U}\cdot\nabla  \approx \vec{U}_\bot\cdot\nabla_\bot \approx \vec{U}_\bot^{apr}\cdot\nabla_\bot$. Under the scaling (\ref{drift_scaling}) and (\ref{weak_z_dep}) we can also neglect the right-hand-side of Eq. (\ref{parallel_momentum_eq}). Conversely, on the right-hand-side of Eq. (\ref{MHD_momentum}) we neglect the second-order terms coming from the polarizaton, charge separation, displacement current and the curvature of the magnetic field. Only the leading part of the last term remains and the equation simplifies to
\begin{equation}\label{pres_bal}
\nabla_\bot\left(\frac{c^2 \epsilon_0 B^2}{2} + p_{e_\bot} + p_{i_\bot}\right) = - \vec{b}\times \left(\frac{p_{i_\bot}}{2\Omega_i} \, \nabla_\bot^2 \vec{U}_{i_\bot} + \frac{p_{e_\bot}}{2\Omega_e} \, \nabla_\bot^2 \vec{U}_{e_\bot}\right),
\end{equation}
which reduces to the equation of pressure balance when the Larmor radius corrections can be neglected.
Now, using
\begin{eqnarray}
\label{div_U_E_U_D}
&& \nabla\cdot\left(\vec{U}_E + \vec{U}_D\right) {\approx }- \left[\frac{\partial}{\partial t} + \left(\vec{U}_E + \vec{U}_D\right)\cdot\nabla_\bot\right]\log B - \left(\vec{U}_E + \vec{U}_D\right)\cdot\left(\vec{b}\cdot\nabla\right) \vec{b}
{- \frac{1}{q n^2 B}\left(\vec{b}\times\nabla p_\bot\right)\cdot\nabla n,}\\
\label{div_U_A}
&& \nabla\cdot\vec{U}_A \approx \left(\vec{b}\cdot\nabla\right)\left[\frac{p_{\Vert_0} - p_{\bot_0}}{q n_0 B_0} \;\; \nabla\cdot\left(\vec{e}_z\times\vec{b}\right)\right],
\end{eqnarray}
and the 
expression (\ref{U_pi}), (\ref{curvature2}) for $\nabla\cdot\vec{U}_\pi$, {we evaluate continuity and parallel momentum equations to leading order in $\epsilon$ setting $\vec{U}\cdot\nabla\approx \vec{U}_\bot^{apr}\cdot\nabla_\bot$ and $\vec{U}_\bot\approx \vec{U}_\bot^{apr}$, and dropping the last term on the right hand side of Eq. (\ref{div_U_E_U_D}) that vanishes if $p_\bot$ is an arbitrary function of $n$, described by Eq. (\ref{pressure}) when $\gamma_\bot=\gamma_\bot(n)$. We obtain:}
\begin{eqnarray}
\nonumber
&&
\hspace{-.4cm}
\left(\frac{\partial}{\partial t} + \vec{U}_\bot^{apr}\cdot\nabla_\bot\right)\left(\log n - \log B\right) + \left(\vec{b}\cdot\nabla\right)\left[U_\Vert + \frac{p_{\Vert_0} - p_{\bot_0}}{q n_0 B_0} \;\; \nabla\cdot\left(\vec{e}_z\times\vec{b}\right)\right] + \\
&&
\frac{1}{\Omega_0}\nabla_\bot\cdot\left\{\left[\frac{\partial}{\partial t} + \left(\vec{U}_\bot^{apr} + \vec{U}_B - \vec{U}_D\right)\cdot\nabla_\bot\right]\vec{e}_z\times\vec{U}_\bot^{apr}\right\}
{+ \rho_{L}^2\left(\nabla_\bot^2\vec{b}\cdot\nabla U_{\Vert}-\vec{b}\cdot\nabla \nabla_\bot^2 U_{\Vert}\right)}
= 0,
\label{new_continuity}
\\
&&
\nonumber
\hspace{-.4cm}
\left[\frac{\partial}{\partial t} + \left(\vec{U}_\bot^{apr} + \vec{U}_B - \vec{U}_D\right)\cdot\nabla\right]U_\Vert =
\\
&&
\frac{q}{m}\left(\vec{b}\cdot\vec{E}
- \rho_{L}^2\, \frac{\partial{\vec{U}_\bot^{apr}}}{\partial x_{\alpha_\bot}}\cdot\nabla_\bot\frac{\partial A_z}{\partial x_{\alpha_\bot}}
\right) - \frac{1}{n_0 m}\left(\vec{b}\cdot\nabla\right)\left[p_\Vert - \left(p_{\Vert_0} - p_{\bot_0}\right)\frac{B}{B_0} + \frac{2 p_{\Vert_0}-p_{\bot_0}}{\Omega_0} \; \nabla_\bot\cdot\left(\vec{b} \times \vec{U}_\bot^{apr}\right)\right].
\label{parallel_momentum_aprox}
\end{eqnarray}
It is convenient to subtract the electron and ion continuity equations, i.e. to use the continuity equation for electric charge, which after neglecting the electrons' polarization and FLR effects yields
\begin{eqnarray}
\nonumber
&&
\left(\vec{U}_{i_\bot}^{apr} - \vec{U}_{e_\bot}^{apr}\right)\cdot\nabla_\bot\left(\log n - \log B\right) + \left(\vec{b}\cdot\nabla\right)\left[U_{i_\Vert} - U_{e_\Vert} + \frac{p_{i_{\Vert 0}} - p_{i_{\bot 0}} + p_{e_{\Vert 0}} - p_{e_{\bot 0}}}{e n_0 B_0} \;\; \nabla\cdot\left(\vec{e}_z\times\vec{b}\right)\right] +
\\
&& \frac{1}{\Omega_{i_0}}\nabla_\bot\cdot\left\{\left[\frac{\partial}{\partial t} + \left(\vec{U}_{i_\bot}^{apr} + \vec{U}_{i_B} - \vec{U}_{i_D}\right)\cdot\nabla_\bot\right]\vec{e}_z\times\vec{U}_{i_\bot}^{apr}\right\}
{+ \rho_{L i}^2 \left(\nabla_\bot^2\vec{b}\cdot\nabla U_{i_\Vert} - \vec{b}\cdot\nabla \nabla_\bot^2 U_{i_\Vert}\right)}
= 0,
\label{charge_continuity}
\end{eqnarray}
Similarly, multiplying the electron and ion momentum equations respectively by $m_e$ and $m_i$ and adding, we obtain the momentum equation for the plasma fluid that in the limit of massless electrons $m_e\to 0$ has the form
\begin{eqnarray}
&&
\nonumber
\hspace{-.4cm}
\left[\frac{\partial}{\partial t} + \left(\vec{U}_{i_\bot}^{apr} + \vec{U}_{i_B} - \vec{U}_{i_D} \right)\cdot\nabla\right]U_{i_\Vert} =
-\rho_{L i}^2\, \frac{q_i}{m_i}\frac{\partial{\vec{U}_{i_\bot}^{apr}}}{\partial x_{\alpha_\bot}}\cdot\nabla_\bot\frac{\partial A_z}{\partial x_{\alpha_\bot}} -
\\
&&
\frac{1}{n_0 m_i}\left(\vec{b}\cdot\nabla\right)\left[p_{i_\Vert} + p_{e_\Vert} - \left(p_{i_{\Vert 0}} - p_{i_{\bot 0}}  + p_{e_{\Vert 0}} - p_{e_{\bot 0}} \right)\frac{B}{B_0} + \frac{2 p_{i_{\Vert 0}}-p_{i_{\bot 0}}}{\Omega_{i_0}} \; \nabla_\bot\cdot\left(\vec{b} \times \vec{U}_{i_\bot}^{apr}\right)\right].
\label{parallel_Plasma_momentum}
\end{eqnarray}
Within the adopted drift- and small-but-finite FLR scalings, Eqs. (\ref{drift_scaling}), (\ref{weak_z_dep}), and taking that the compressional perturbation of the magnetic field is of the same order as, or smaller than the torsional component, the electromagnetic field can be expressed in terms of the electrostatic potential and of the $z$-components of the vector potential and magnetic field, $\phi$, $A_z$, and $\delta B_z$, viz.
\begin{equation}
\label{E_B_field}\vec{E} = -\nabla\phi - \frac{\partial}{\partial t}\left(\vec{e}_z A_z + \vec{e}_z\times\nabla_\bot \; \nabla_\bot^{-2} \delta B_z\right),
\quad\quad
\vec{B} = \vec{e}_z \left(B_0 + \delta B_z\right) - \vec{e}_z\times\nabla_\bot A_z,
\end{equation}
which yields the following expressions, accurate to first order:
\begin{eqnarray}
\label{unit_vectors}
&&\vec{b} = \vec{e}_z - \frac{1}{B_0} \; \vec{e}_z\times\nabla_\bot A_z,
\quad\quad
\vec{e}_l = \vec{e}_x - \frac{\vec{b}}{B_0} \frac{\partial A_z}{\partial y},
\quad\quad
\vec{e}_m = \vec{e}_y + \frac{\vec{b}}{B_0} \frac{\partial A_z}{\partial x},
\\
\label{pppp}
&&
B = |\vec{B}| = B_0 + \delta B_z,
\quad\quad
\vec{b}\cdot\vec{E} = -\vec{b}\cdot\nabla\phi - \frac{\partial A_z}{\partial t},
\quad\quad
\vec{b}\times\vec{E} = -\vec{e}_z\times\nabla_\bot\phi,
\end{eqnarray}
We conveniently rewrite our basic equations in a dimensionless form using the following scaled variables and parameters
\begin{eqnarray}
\nonumber &&
t'= \omega\, t , \quad x' =  k\, x, \quad y' = k\, y, \quad z' = (\omega/c_A)\, z,
\quad
\phi'= \frac{k^2 \phi}{B_0 \omega}, \quad A_z' = \frac{k^2 c_A A_z}{B_0 \omega}, \quad B_z' = \frac{\Omega_{i_0}}{\omega} \frac{\delta B_z}{B_0}, \quad n' = \frac{\Omega_{i_0}}{\omega} \frac{\delta n}{n_0}, 
\\ &&
U_\Vert' = \frac{\omega_{p i}}{\omega} \frac{U_\Vert}{c}, \quad p' =\frac{k^2}{\Omega_{i_0}\omega}\frac{\delta p}{n_0 m_i}
, \quad
\beta = \frac{2 p_0}{c^2 \epsilon_0 B_0^2}, \quad
d_e^{\,\prime} = \frac{c\, k}{\omega_{pe}}, \quad d_i^{\,\prime} = \frac{c\, k}{\omega_{pi}}, \quad
\rho_i' = k\, \rho_{L i} =
d_i^{\,\prime} \, \sqrt{\frac{\beta_{i_\bot}}{2}},
\label{dimensionless_variables}
\end{eqnarray}
where $k$ and $\omega$ are some characteristic wavenumber and characteristic frequency, i.e. the inverse characteristic spatial and temporal scales, such as the width of the structure $r_0$ and the transit time $r_0/u_\bot$, $u_\bot$ being the speed of its propagation in the plasma frame transversely to the magnetic field. The normalization through $k, \omega$ is used for convenience and does not infer any presence of wave phenomena. Other notations are standard, $\omega_{p i} = \sqrt{n_0 e^2/m_i\epsilon_0}$ is the ion plasma frequency and $c_A = c \, \Omega_{i_0}/\omega_{p i}$ is the Alfv\'{e}n speed. Note that the dimensionless parallel velocity $U'_\Vert$, the pressure $p'$, and the parameter $\beta$ involve either the electrons or the ions, and the two latter quantities also the perpendicular or the parallel components, which is denoted below by the appropriate combination of the subscripts $e, i, \bot$, and $\Vert$.
The dimensionless versions of the charge continuity, Eq. (\ref{charge_continuity}), the electron continuity and parallel momentum equations, Eqs. (\ref{new_continuity}) and (\ref{parallel_momentum_aprox}), and of the parallel and  perpendicular fluid momentum equations (\ref{parallel_Plasma_momentum}) and  (\ref{pres_bal}), can be written as follows (for simplicity, here and in the rest of the paper, we omit the primes):
\begin{eqnarray} \nonumber
&& \hspace{-.5cm}
\left[\vec{e}_z\times\nabla_\bot \left(\frac{p_{i_\bot}-\rho_i^2\nabla_\bot^2\phi/2}{1 + \rho_i^2\nabla_\bot^2/2} + p_{e_\bot}\right)\right] \cdot\nabla_\bot\left(n - B_z\right) - \left[\frac{\partial}{\partial z} - \left(\vec{e}_z\times\nabla_\bot A_z\right)\cdot\nabla_\bot\right]\left(1 - \frac{\beta_{i_\Vert} - \beta_{i_\bot} + \beta_{e_\Vert} - \beta_{e_\bot}}{2}\right) \nabla_\bot^2 A_z
\\
\nonumber
&& \hspace{1.5cm}
-\nabla_\bot\cdot\left(\left\{\frac{\partial}{\partial t} + \left[\vec{e}_z\times\nabla_\bot\left(\frac{\phi - \rho_i^2 \nabla_\bot^2 p_{i_\bot}/2}{1 + \rho_i^2\nabla_\bot^2/2} + \frac{\beta_{i_\bot}}{2} \; d_i^2 B_z\right)\right]\cdot\nabla_\bot\right\}\cdot\nabla_\bot \; \frac{\phi + p_{i_\bot}}{1 + \rho_i^2\nabla_\bot^2/2}\right)
\\
&&
\hspace{1.5cm}
{- \rho_i^2\left(\vec{e}_z\times\nabla_\bot \nabla_\bot^2 A_z\right) \cdot\nabla_\bot U_{i_\Vert} -
\rho_i^2 \left[\frac{\partial}{\partial z}- \left(\vec{e}_z\times\nabla_\bot A_z\right) \cdot\nabla_\bot\right]\nabla_\bot^2 U_{i_\Vert}}= 0,
\label{charge_cont_DL}
\end{eqnarray}
\begin{eqnarray}
&&
\hspace{-2.1cm}
\left\{\frac{\partial}{\partial t} + \left[\vec{e}_z\times\nabla_\bot \left(\phi - p_{e_\bot}\right)\right]\cdot\nabla_\bot\right\}\left(n - B_z\right) +
\left[\frac{\partial}{\partial z} - \left(\vec{e}_z\times\nabla_\bot A_z\right)\cdot\nabla_\bot\right]
\left(U_{e_\Vert} - \frac{\beta_{e_\Vert} - \beta_{e_\bot}}{2} \nabla_\bot^2 A_z \right) = 0,
\label{elec_cont_DL}
\end{eqnarray}
\begin{eqnarray}
&&
\hspace{-7.5cm}
\left[\frac{\partial}{\partial z} - \left(\vec{e}_z\times\nabla_\bot A_z\right)\cdot\nabla_\bot\right]
\left(\phi -p_{e_\Vert} + \frac{\beta_{e_\Vert} - \beta_{e_\bot}}{2} \; d_i^2 B_z\right) +
\frac{\partial A_z}{\partial t}
=0,
\label{par_mom_elec_DL}
\end{eqnarray}
\begin{eqnarray}\nonumber
&&\hspace{-.4cm}
\left\{\frac{\partial}{\partial t} + \left[\vec{e}_z\times\nabla_\bot\left(\frac{\phi - \rho_i^2 \nabla_\bot^2 p_{i_\bot}/2}{1+ \rho_i^2\nabla_\bot^2/2} + \frac{\beta_{i_\bot}}{2} \; d_i^2 B_z\right)\right]\cdot\nabla_\bot\right\} \, d_i^2 U_{i_\Vert} =
\\
\nonumber
&&
\hspace{1.5cm}
- \left[\frac{\partial}{\partial z} - \left(\vec{e}_z\times\nabla_\bot A_z\right)\cdot\nabla_\bot\right]
\left[p_{i_\Vert}+ p_{e_\Vert} - \frac{\beta_{i_\Vert} - \beta_{i_\bot}+ \beta_{e_\Vert} - \beta_{e_\bot}}{2} \; d_i^2 B_z
- \left(\frac{2\beta_{i_\Vert}}{\beta_{i_\bot}}-1\right) \frac{\rho_i^2 \nabla_\bot^2\left(\phi + p_{i_\bot}\right)}{1 + \rho_i^2\nabla_\bot^2/2}\right]
\\
&&
\hspace{1.5cm}
- \rho_i^2\left(\vec{e}_z\times\nabla_\bot \; \frac{\partial}{\partial x_{\alpha_\bot}}\frac{\phi +
p_{i_\bot}}{1 + \rho_i^2\nabla_\bot^2/2}\right)\cdot\nabla\frac{\partial A_z}{\partial x_{\alpha_\bot}},
\label{par_mom_ion_DL}
\end{eqnarray}
\begin{eqnarray}
&&\hspace{-10cm}
\label{pres_bal_DL}
d_i^2 B_z + p_{i_\bot} + p_{e_\bot} - \frac{\rho_i^2\nabla_\bot^2\left(\phi + p_{i_\bot}\right)/2}{1 + \rho_i^2\nabla_\bot^2/2} = 0,
\end{eqnarray}
where we neglected the charge separation ($n_e = n_i = n$) and the displacement current ($U'_{e_\Vert} - U'_{i_\Vert} = {\nabla'_\bot}^{\!\! 2} A'_z$), and we considered the electrons to be massless (i.e. we took $m_e\to 0$ and accordingly $\omega/\Omega_e = \rho'_e = d'_e = 0$).

Pressure perturbations are expressed in terms of density via dimensionless versions of the equations of state (\ref{pressure}) viz.
\begin{equation}\label{eq_state_DL}
p_{j\,\zeta}' = \gamma_{j\, \zeta} \, \left({\beta_{j\, \zeta}}/{2}\right) \; {d_i^{\, \prime}}^2 n', \quad {\rm where} \quad j = e, i \quad {\rm and} \quad \zeta = \Vert, \bot.
\end{equation}
It is worth noting that the acoustic perturbations, that are associated with the parallel plasma motion described by the momentum equation (\ref{par_mom_ion_DL}), are coupled with the rest of the system only through the last terms in Eq. (\ref{charge_cont_DL}), the first of which has come from the divergence of the curvature correction to the FLR drift, $\nabla\cdot\delta\vec{U}_{i\pi}$. It will be shown below that in the regimes of our interest these terms are small and the acoustic perturbations are fully decoupled.

Eqs. (\ref{charge_cont_DL})--(\ref{par_mom_elec_DL}) and (\ref{pres_bal_DL}) constitute our basic set of equations for the functions $\phi, A_z, B_z$, and $n$. In a plasma with arbitrary values of $\beta_{j\, \zeta}$, these equations describe variations that are slowly varying both in time and along the magnetic field lines, but have an arbitrary spatial scale perpendicular to the magnetic field.

We readily note that all nonlinear terms in our Eqs. (\ref{charge_cont_DL})-(\ref{par_mom_ion_DL}) have the form of mixed products that vanish when the solution is one-dimensional, either in Cartesian or in cylindrical coordinates. As a consequence, there exist trivial solutions that are independent on $z$ and stationary in time, $\partial/\partial z = \partial/\partial t = 0$, such as current sheets (planar discontinuities) and localized cylindrically symmetric solutions that may have arbitrary radial dependencies. The latter have the form of Petviashvili-Pokhotelov monopoles and are constrained only by the equations of pressure balance (\ref{pres_bal_DL}) and state (\ref{eq_state_DL}). The monopoles come in two distinct varieties, as the electrostatic and magnetostatic convective cells (the latter are often referred to as the current filaments or force-free currents) \citep{1983PhFl...26.1388S,1984PhRvL..53.2559S}.
Force free currents, in the form of filaments parallel to the magnetic field lines \citep{1994ppit.book.....S,2015PrPh...11b.167S}, are often encountered
in space plasmas - in the solar corona, planetary magnetospheres, solar wind, etc. and they are thought to have a sufficiently long lifetime to be carried by the solar wind for several AU. The evolution of the electrostatic cells has been extensively studied both numerically and experimentally, see e.g.  Refs. \citep{1980PhLA...80...23L,1984PhRvL..52.2148P}. The experiments in quiescent plasma, as well as the water tank experiments  \citep{2003PhFl...15.1033B}, in which the perturbations are described by 2-d (two dimensional) model equations that are similar to those in the plasma but include also a small but finite viscosity, revealed that stationary monopoles either disperse or slowly transform into propagating dipolar or rotating tripolar vortices, depending on the amount of shielding included in the radial profile of the initial state.
%
%
Finding a general solution of the above equations is a formidable task because of their complexity, but suitable
particular solutions have been found in a number of relevant cases. Thus, in our earlier papers we studied in detail the electron-scale nonlinear structures in plasmas with cold electrons and very cold ions, $\beta_{e_\bot}\ll 1$ and $\beta_{i_\bot}\ll m_e/m_i$ on the MHD temporal scale \citep{1987JPlPh..37...81J,1994PhPl....1.2614J}. For various travelling solutions in plasma regimes in which all diamagnetic drift- and FLR effects are negligible see also \citep{1992swpa.book.....P,2006JGRA..11112208A}. An extensive study of drift-Alfv\'en vortices in the regime $\beta\ll 1$ was presented by \citet{1999PhPl....6..713K}. In the warm plasma, mostly electron scale structures were considered, such as whistler-frequency perturbations in the regime $\beta_{e_\bot}\gtrsim 1$ {with immobile ions} \citep{2015PhyS...90h8002J}.

In the present paper, firstly we present a general solution that is slightly bigger that the ion-scale (i.e. we keep the leading FLR terms) in a plasma with, $\beta_{i_\bot} \sim \beta_{e_\bot}\sim 1$, described by the Kadomtsev-Pogutse-Strauss' equations of reduced MHD \citep{1973ZhETF..65..575K,1974ZhETF...66.2056K,
1976PhFl...19..134S,1977PhFl...20.1354S} and identified as a Chaplygin vortex \citep{Chaplygin}.  A class of stationary, magnetic field aligned monopoles can be derived from such travelling solutions in the limit of zero inclination and zero propagation speed. {Due their smooth transition to the linear response at large distances, these can be expected to be more stable than the monopoles (magnetostatic cells) with arbitrary radial dependencies}. Additionally, in a plasma with $\beta_{i_\bot} \sim \beta_{e_\bot}\lesssim 1$, we find two kinds of dipolar Larichev\& Reznik-type \citep{Larichev_Reznik} ion-scale vortices, that can be identified as the high-$\beta$ shear-Alfv\'{e}n and kinetic slow nonlinear modes, respectively. The corresponding quasi-monopoles contain thin {(electron scale)} current sheets and/or charge layers at the core edge, and are probably less stable, particularly those in the kinetic slow mode branch. Like all Larichev\&Reznik-type solutions, inside the core the vortices are fundamentally nonlinear structures, while outside of the core they are described by the corresponding linear equations and thus follow (at least asymptotically) the linear dispersion relation in cylindrical coordinates {of shear Alf\'{e}n and kinetic slow magnetosonic modes}. Thus, they can also be regarded as the generalization and the regularization of the singular point vortex model that is broadly used in the study of the 2-d turbulence in incompressible fluids and plasmas \citep{Kirchhof,2016PlPhR..42..523K}, see also the collection of theoretical and numerical works \citep{1989mavd.proc.....C}. Such nonlinear regularization will also broaden the envisaged spectrum of an agglomerate of point vortices.

\section{Vortex solutions} 
\label{Traveling}
\subsection{Solution sufficiently bigger than the ion Larmor radius}\label{large_scale}

One can easily verify that our model equations (\ref{charge_cont_DL})-(\ref{par_mom_elec_DL})
reduce to the Kadomtsev-Pogutse-Strauss' equations of reduced MHD when all terms arising from the diamagnetic drift and FLR effects can be neglected. Using our notation introduced in Eq. (\ref{dimensionless_variables}), this condition is expressed as $\rho_i' = d_i^{\, \prime}\, \sqrt{\beta_{i_\bot}/2}\ll 1$, which is obviously satisfied when either the plasma $\beta$ is sufficiently small or the characteristic scale largely exceeds the ion inertial length  $c/\omega_{p i}$, i.e. for $d_i'\ll 1$. However, it is worth noting that the dimensionless ion Larmor radius $\rho_i'$ can be  sufficiently small also in plasmas with a modest value of $\beta$, even if the size of the solution is not much bigger than the ion inertial length, i.e. $d_i^{\, \prime}$ is not too small (e.g. when $d_i^{\, \prime}\lesssim 0.4$). For example, in the magnetosheath region
with $\beta_{i_\bot} =
0.5$, Cluster mission detected structures whose spatial scale was $r_0\simeq 500 \, \rm km$ \citep{2004JGRA..109.5207A}, more than ten times the ion inertial depth $c/\omega_{pi} \simeq 40 \, \rm km$. Under such conditions, we have $d_i^{\,\prime} \lesssim 0.1$ and a very small value of ${\rho_i'}$, viz.
${\rho_i'}^2 = (\beta_{i_\bot}/2) \, {d_i^{\, \prime}}^2 \approx .0025 \ll 1$. {Recently, quasi-monopolar Alfv\'{e}n vortex with the transverse radius of $\sim 10$ proton gyroscales, was identified and studied in detail \citep{2019ApJ...871L..22W} using the data collected by the Magnetospheric Multiscale mission (MMS) in the Earth's turbulent magnetosheath.
}

Obviously, for perturbations that comply with the small ion Larmor radius scaling
\begin{equation}\label{novo_skaliranje}
\rho_i'= \left(k/\Omega_i\right)\sqrt{T_{i_\bot}/m_i} \sim\epsilon,
\end{equation}
where $\epsilon$ is the small parameter introduced in Eqs. (\ref{drift_scaling}) and (\ref{weak_z_dep}), we can set ${\rho_i'}^2 
\to 0$. Then, Eq. (\ref{eq_state_DL}) yields $p'_{j \zeta} \to 0$ even if the dimensionless density perturbation is finite,  $n'\sim 1$. The latter implies also $\vec{E}_\Vert = 0$, which excludes both kinetic-Alfv\'{e}n and magnetosonic waves from our description, see Eq. (\ref{par_mom_long_scale}). Under these conditions, within the drift- and weak $z$-dependence scalings,  Eqs. (\ref{drift_scaling}) and (\ref{weak_z_dep}), and in the massless electron limit (i.e. for perturbations whose spatial scale is much bigger than the electron inertial length, viz. $d_e'\to 0$), our Eqs. (\ref{charge_cont_DL}) and (\ref{par_mom_elec_DL}) decouple from the rest of the system. They possess the form of the  standard Kadomtsev-Pogutse-Strauss' reduced MHD system for the potentials $\phi$ and $A_z$, including the effects of anisotropic temperatures (henceforth, we drop the primes for simplicity)
\begin{eqnarray}
    && \left[\frac{\partial}{\partial t} + \left(\vec{e}_z\times\nabla_\bot \phi\right)\cdot\nabla_\bot\right] \nabla_\bot^2\phi =
    -\left[\frac{\partial}{\partial z} - \left(\vec{e}_z\times\nabla_\bot A_z\right)\cdot\nabla_\bot\right]
    \left(1 - \frac{\beta_{i_\Vert} - \beta_{i_\bot} + \beta_{e_\Vert} - \beta_{e_\bot}}{2}\right) \nabla_\bot^2 A_z,
    \label{plasma_cont_long_scale}
    \\
    && \frac{\partial A_z}{\partial t} + \left[\frac{\partial}{\partial z} - \left(\vec{e}_z\times\nabla_\bot A_z\right)\cdot\nabla_\bot\right]\phi = 0,
    \label{par_mom_long_scale}
    \end{eqnarray}
while {the parallel velocity,} the density and the compressional magnetic field are subsequently determined from:
\begin{eqnarray}\nonumber
&&
{
\left[\frac{\partial}{\partial t} + \left(\vec{e}_z\times\nabla_\bot\phi\right)\cdot\nabla_\bot\right] \, U_{i_\Vert} +
\frac{\beta_{i_\bot}}{2}\left(\vec{e}_z\times\nabla_\bot \; \frac{\partial\phi}{\partial x_{\alpha_\bot}}\right)\cdot\nabla\frac{\partial A_z}{\partial x_{\alpha_\bot}} =
}
\\
&&
\hspace{1.5cm}
{
- \left[\frac{\partial}{\partial z} - \left(\vec{e}_z\times\nabla_\bot A_z\right)\cdot\nabla_\bot\right]
\left(\frac{\beta_{i_\Vert}+ \beta_{e_\Vert}}{2}\; n - \frac{\beta_{i_\Vert} - \beta_{i_\bot}+ \beta_{e_\Vert} - \beta_{e_\bot}}{2} \; B_z
- \frac{2\beta_{i_\Vert}-\beta_{i_\bot}}{2}\; \nabla_\bot^2\phi\right),
}
\label{par_mom_ion_long_scale}
\\
&& \label{el_cont_long_scale}
    \left[\frac{\partial}{\partial t} + \left(\vec{e}_z\times\nabla_\bot\phi\right)\cdot\nabla_\bot\right]\left(n - B_z\right) =
    -\left[\frac{\partial}{\partial z} - \left(\vec{e}_z\times\nabla_\bot A_z\right)\cdot\nabla_\bot\right]
    \left[{U_{i_\Vert}+}\left(1 - \frac{\beta_{e_\Vert} - \beta_{e_\bot}}{2}\right)
    \nabla_\bot^2 A_z\right],
    \\
    && \label{pres_bal_long_scale}
    B_z +
\frac{{\beta_{i_\bot} + \beta_{e_\bot}}}{2} \; n -  \frac{\beta_{i_\bot}}{4} \; \nabla_\bot^2 \phi = 0.
    \end{eqnarray}
{Here, in the regime $\rho_i^2\to 0$, we have used $\gamma_{i_\bot}=\gamma_{e_\bot}=1$ as discussed above, see Eq. (\ref{pressure}).}

Note that Eqs. (\ref{plasma_cont_long_scale})-(\ref{pres_bal_long_scale}) have no spatial scales, i.e. the characteristic wavenumber and frequency, $k$ and $\omega$  introduced in Eq. (\ref{dimensionless_variables}) (i.e. the width of the structure, $r_0$, and its speed in the plasma frame, $u_\bot$) are arbitrary.
%
%
Multiplying Eqs. (\ref{plasma_cont_long_scale}) and (\ref{par_mom_long_scale}) by $\phi$ and $[ 1 - (1/2)({\beta_{i_\Vert} - \beta_{i_\bot} + \beta_{e_\Vert} - \beta_{e_\bot}})]\, \nabla_\bot^2 A_z$, respectively, using the identities $(\vec{e}_z\times\nabla f)\cdot\nabla g=\nabla\cdot(g\;\vec{e}_z\times\nabla f)$ and $\Phi\, \nabla_\bot^2\partial\Phi/\partial t = \nabla\cdot(\Phi \, \nabla_\bot\partial\Phi/\partial t)-(\partial/\partial t)(\nabla_\bot\Phi)^2/2$, and integrating for entire space (provided $\phi=A_z=0$ at infinity), we readily obtain that the energy $W$ is conserved, $\partial W/\partial t = 0$, where
\begin{equation}\label{energy_long_scale}
W = \left({1}/{2}\right)\int_{-\infty}^\infty\int_{-\infty}^\infty\int_{-\infty}^\infty dx \, dy\, dz \left\{\left(\nabla_\bot\phi\right)^2 +
\left[1 - \left(1/2\right)\left({\beta_{i_\Vert} - \beta_{i_\bot} + \beta_{e_\Vert} - \beta_{e_\bot}}\right)\right]\left(\nabla_\bot A_z\right)^2\right\}.
\end{equation}
{Our equations (\ref{plasma_cont_long_scale}) and (\ref{par_mom_long_scale}) are identical (apart from constant factors that come from the temperature anisotropy) to} the Kadomtsev-Pogutse-Strauss' system that has been studied in details in the literature, see Ref. \citep{1992swpa.book.....P}  and references therein. Following the standard procedure of \citet{Larichev_Reznik} and of more recent works \citep{1999PhPl....6..713K}, we seek its solution that is travelling with the (non scaled) velocity $\vec{u}_\bot = \vec{e}_y \, k/\omega$ and is tilted to the $z$-axis by a small angle $\theta = \omega/k u_z$, implying that the solution depends only on the dimensionless variables $x'$ and $y' + (c_A/u_z)\, z' - t'$. Then, using $\partial/\partial t' = -\partial/\partial y'$ and $\partial/\partial z' = (c_A/u_z) \, \partial/\partial y'$, the parallel electron momentum Eq. (\ref{par_mom_long_scale}) is readily solved as
\begin{equation}\label{no_parallel_E}
\phi = \left({u_z}/{c_A}\right)\; A_z,
\end{equation}
indicating that the parallel electric field is equal to zero, $E_\Vert=0$, short circuited by the 
massless electrons. As a consequence, $\delta\vec{B}_\bot$ is aligned with $\vec{U}_\bot$. Substituting the above into Eq. (\ref{plasma_cont_long_scale}) yields a simple 2-D Euler equation
\begin{equation}\label{integr_cont_eqs_Strauss}
\left[\vec{e}_z\times\nabla_\bot\left(\phi-x\right)\right]\cdot\nabla_\bot\nabla_\bot^2 \phi = 0
\quad \Rightarrow \quad
\nabla_\bot^2\phi = G\left(\phi-x\right).
\end{equation}
Here, $G$ is an arbitrary function of its argument that is adopted here to be part-by-part linear \citep{Larichev_Reznik,1999PhPl....6..713K}, viz.
\begin{equation}\label{part-by-part}
G(\xi)=(\xi - \xi_0)\, G_1,
\end{equation}
where the parameter $\xi_0$ and the slope $G_1$ take different constant values $\xi_0^{in}, G_1^{in}$ and $\xi_0^{out}, G_1^{out}$ inside and outside, respectively, of a moving circle in the $x,y$ plane whose radius is $r_0$ (usually referred to as the vortex core). For a spatially localized solution, we have $G_1^{out} = 0$, which implies that {both the vorticity $\nabla_\bot^2\phi$ and the parallel current $-\nabla_\bot^2 A_z$} are localized inside the vortex core, while $G_1^{in}$ will be determined from the smoothness {of the potentials $\phi$ and $A_z$ (i.e. from the absence of the surface charges and surface currents along $\vec{B}_0$)} at the edge of the vortex core.

Eq. (\ref{integr_cont_eqs_Strauss}) separates variables in cylindrical coordinates. The solution is easily written in terms of the Bessel functions,
\begin{equation}\label{Petv_Pokh}
\phi(r,\varphi) = \sum_k [\alpha_k J_k(r)+\beta_k Y_k(r)] \exp(i k \varphi),
\end{equation}
where $r =\{x^2 + [y+(c_A/u_z)\, z - t]^2\}^\frac{1}{2}$, $\varphi = \arctan\{[y+(c_A/u_z)\, z - t]/x\}$, and $J_k$ and $Y_k$ are the Bessel functions of the k-th order and of the first and second kind, respectively. The constants of integration $\alpha_k$ and $\beta_k$ are determined from the finiteness of the solution at $r=0$ and $r\to\infty$, and from the physical conditions of continuity and smoothness at the core edge $r=r_0$ of the potential $\phi$. The ensuing solution takes the form of a \citet{Chaplygin} vortex, constructed more than a century ago as the traveling solution of a 2-D Euler equation for the incompressible flow in ordinary fluids. It consists of a circularly symmetric "rider" that is appropriately superimposed on a Lamb dipole that also provides its propagation, viz.
\begin{equation}
\label{celo_resenje_ion_scale}
\phi\left(r, \varphi\right) = \left(u_z/c_A\right)\; A_z\left(r, \varphi\right) =
\left\{\begin{array}{ll}
   \cos\varphi \; \left(r_0^2/r\right), & \hbox{$r\geq r_0$,} \\
   \cos\varphi \; \left\{r - (2 r_0/j_1) \; [J_1\left(j_1 \; r/r_0\right)/J_1'\left(j_1\right)]\right\}
   + \psi_0\left[J_0\left(j_1 \; r/r_0\right) - J_0\left(j_1\right)\right] ,  & \hbox{$r<r_0$,}
\end{array} \right. ,
\end{equation}
where
$j_k$ is one of the zeros of the Bessel function $J_{1}$, viz. $J_{1}(j_k) = 0$, and the amplitude $\psi_0$ of the monopole component is arbitrary. The solution (\ref{celo_resenje_ion_scale}) does not possess a characteristic spatial scale and the radius $r_0$ is arbitrary. An identical Alfv\'{e}n vortex was presented by \citet{1992swpa.book.....P}, albeit in Cartesian coordinates.

The corresponding density, compressional magnetic field, and parallel ion velocity are found from Eqs. (\ref{par_mom_ion_long_scale})--(\ref{pres_bal_long_scale}) and using (\ref{no_parallel_E})
, viz.
{
\begin{eqnarray}
\nonumber
n \equiv {\cal N}\; \nabla_\bot^2\phi &=& \left[1+\frac{\beta_{e_\bot}+\beta_{i_\bot}}{2} + \frac{c_A^2}{u_z^2}
\left(\frac{\beta_{e_\bot}+\beta_{i_\bot}}{2}
\frac{\beta_{e_\bot}-\beta_{e_\Vert}+\beta_{i_\bot}-\beta_{i_\Vert}}{2} -\frac{\beta_{e_\Vert}+\beta_{i_\Vert}}{2}\right)\right]^{-1} \times
\\
&&
\left\{\frac{\beta_{i_\bot}}{4} + \frac{c_A^2}{u_z^2}\left[ 1-\frac{\beta_{i_\Vert}}{2} + \frac{\beta_{e_\bot}-\beta_{e_\Vert}+\beta_{i_\bot}-\beta_{i_\Vert}}{2}
\left(1+\frac{\beta_{i_\bot}}{4}\right)\right]\right\}\,\nabla_\bot^2\phi,
\label{n_petv_pokh}
\\
B_z \equiv {\cal B}\;\nabla_\bot^2\phi &=& \left[-\frac{\beta_{i_\bot}+\beta_{e_\bot}}{2}\,{\cal N} + \frac{\beta_{i_\bot}}{4}\right]\,\nabla_\bot^2\phi.
\label{B_petv_pokh}
\\
U_{i_\Vert} \equiv {\cal U}_i\;\nabla_\bot^2\phi &=& \frac{u_z}{c_A} \left[{\cal N}-{\cal B} - \frac{c_A^2}{u_z^2} \left(1-\frac{\beta_{e_\Vert}-\beta_{e_\bot}}{2}\right)\right]\, \nabla_\bot^2\phi.
\label{Ui_petv_pokh}
\end{eqnarray}
}
The above density and magnetic field perturbations are clearly proportional, viz. $B_z/n=\rm constant$. Depending on the plasma parameters and the characteristic velocity $u_z$, they can be either correlated or anti-correlated, $B_z/n>0$ or $<0$ respectively. In a general case, the dependence of ${\rm sign}(B_z/n)$ on plasma parameters is rather complicated and we present only {the results in isothermal plasma, with $\beta_{i_\bot}=\beta_{i_\Vert}=\beta_{e_\bot}=\beta_{e_\Vert}\equiv\beta$, in which the density and magnetic field perturbations are correlated $B_z/n>0$ either for sufficiently large $u_z$, viz. $u_z^2/c_A^2>{\rm max}\,(4-\beta, \; 2-4/\beta)\geqslant 3-\sqrt 5$ or for small parallel phase speeds, $u_z^2/c_A^2<{\rm min}\,(4-\beta,\; 2-4/\beta)\leqslant 3-\sqrt 5$ (the latter case can be realized only for large thermal pressures, $2<\beta<4$ and it involves also acoustic perturbations). Otherwise, $B$ and $n$ are anticorrelated.}

\begin{figure}[htb]
\centering
\includegraphics[width=144mm]{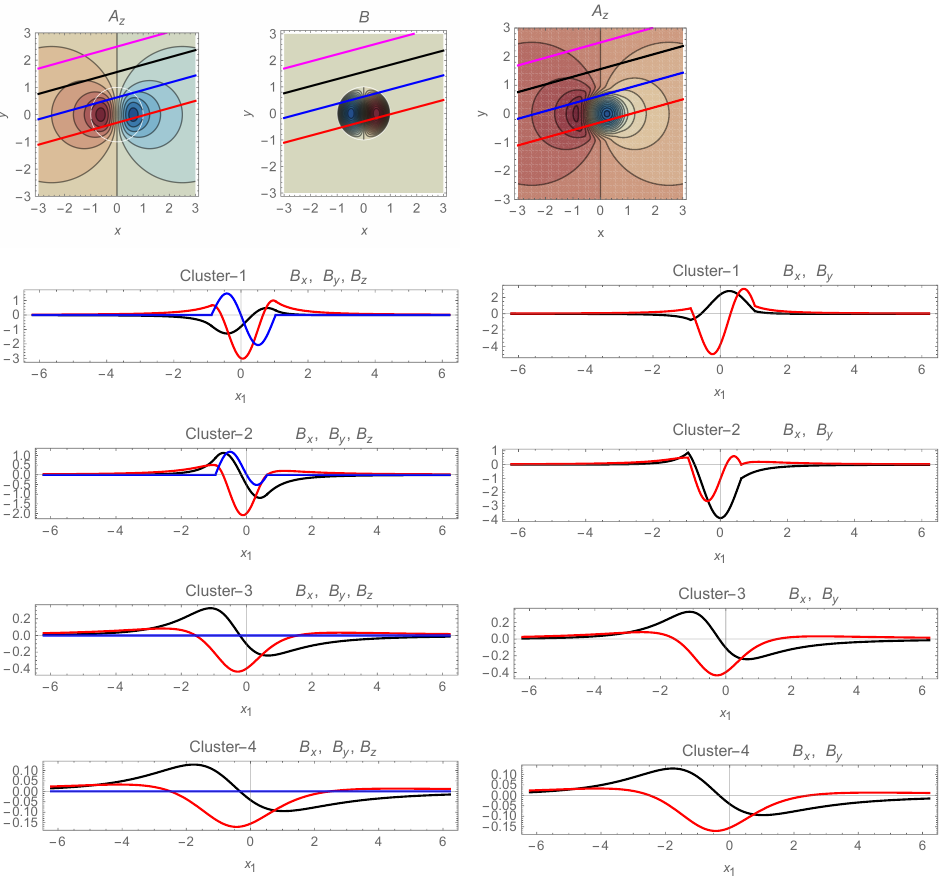}
\caption{Left column: dipolar Lamb vortex with $\psi_0=0$, in which $B_z$ is of the same order as $B_x$ and $B_y$.
\\
Right column: quasi-monopolar Chaplygin vortex with $\psi_0\ne 0$ and $B_z\to 0$.
\\
Top to bottom: \\
Row 1: Contour plots of the vector potential $A_z$ and of the compressional magnetic field $B_z$; in the case of the Chaplygin's vortex the compressional magnetic field is negligibly small. In the figure, vortices propagate in the vertical direction.
\\
Four typical trajectories of the spacecraft, $S$-1, $S$-2, $S$-3, and $S$-4 are displayed as red, blue, black and magenta parallel lines. On the 'black and 'magenta' trajectories no compressional magnetic field $B_z$ is recorded, but the 'detected signals' of the perpendicular magnetic field on 'blue' and 'black' trajectories are of a similar intensity as $B_z$. \\
Rows 2--5: Three components of the dimensionless magnetic field,
as they would be observed by the four spacecrafts on red, blue, black, and magenta trajectories. The coordinate system is rotated with respect to that used in the calculations, so that the $B_x$ component (black line) is now in the direction of the projection of the spacecraft's velocity to the perpendicular plane, $B_y$ (red) is perpendicular both to it and to the magnetic field, and $B_z$ (blue) is parallel to the ambient magnetic field. Vanishing $E_\Vert$ implies that the dimensionless electric field components are given by $E_x=-B_y$, $E_y=B_x$ and $E_z=0$. Density is given by $n/B_z={\rm constant}$. For the  large scale RMHD vortex $n/B_z\ne 1$, see Eqs. (\ref{n_petv_pokh}), (\ref{B_petv_pokh}), and for the ion-scale vortex $n/B_z= 1$. Normalizations are defined in Eq. (\ref{dimensionless_variables}), with $k=1/r_0' = 1$ (i.e. lengths are normalized to $r_0$) and $\psi_0 = 1.75$. (color online) }  \label{slika1}
\end{figure}

The large-scale vortex presented in Eqs. (\ref{celo_resenje_ion_scale})-(\ref{Ui_petv_pokh}) is displayed in Fig. \ref{slika1}.
Note that the vorticity $\nabla_\bot^2\phi$ accociated with the monopolar component of the solution (\ref{celo_resenje_ion_scale}) has a finite jump at $r=r_0$ and that the function $G$ is discontinuous for the corresponding value of its argument. Such discontinuity is permitted, since it does not yield a singularity of the vector-product in Eq. (\ref{integr_cont_eqs_Strauss}). At the core edge, the latter comprises the product of the derivatives in the directions parallel and perpendicular to the isoline $r=r_0$. As the derivative of a function along its isoline is always zero, the corresponding product remains finite even if the perpendicular derivative of $\nabla_\bot^2\phi$ is infinite. In such a case, the vector-product 
acquires an isolated point and remains a continuous function elsewhere.

A stationary, nonpropagating monopole aligned with the background magnetic field is readily obtained in the limit $u_\bot\to 0$ from the moving, tilted, quasi-monopole (\ref{celo_resenje_ion_scale}), whose monopolar component has an arbitrary amplitude and the amplitude of the dipole is proportional to the propagation velocity $u_\bot=\omega/k$. Here $1/\omega$ and $1/k$ are the characteristic temporal and spatial scales introduced in Eq. (\ref{dimensionless_variables}).

The stability of propagating dipoles and quasimonopoles, and of stationary monopoles was studied in the experiments in water tanks \citep{2003PhFl...15.1033B}, in which the perturbations are also described by the 2-d Euler equation (\ref{integr_cont_eqs_Strauss}). However, they lack any 3-D properties but include an additional small but finite viscosity of the fluid, not involved in our analysis. In these experiments, a dipole [solution with $\psi_0=0$ in Eq. (\ref{celo_resenje_ion_scale})] appears to be remarkably stable and it can easily survive collisions with other dipoles \citep{1989Natur.340..212V}. Propagating quasi-monopoles, described by Eq. (\ref{celo_resenje_ion_scale}) when $\psi_0\ne 0$, are found to travel over a distance that is an order of magnitude bigger than its diameter, that has been suggested also by the weak nonlinear theory and numerical simulations \citep{1998JPO....28...22S}, as well as by experiments in non-rotating water tanks with rectangular shape \citep{1999JPO....29.2741V}. In these works, the moving quasi-monopoles were much more stable than the stationary monopoles, that were found to either disperse or slowly transform into dipolar or tripolar vortices depending on their initial profile, i.e. on the amount of shielding in the initial state.

The properties of our theoretical solution (\ref{celo_resenje_ion_scale})--(\ref{Ui_petv_pokh}) (spatial and temporal scales, magnetic and electric field structures, temperature variations and parallel flows, compresibility etc.) closely mimic those of the Alfv\'{e}n vortex \citep{2019ApJ...871L..22W} identified in the high resolution data collected by the MMS mission in the Earth's turbulent magnetosheath.

\subsection{Approaching ion scales}\label{ion_scales}

When the plasma $\beta$ is close to, or or bigger than, unity the dimensionless ion Larmor radius becomes comparable and even bigger than the dimensionless ion inertial length. Then, the anisotropic reduced MHD equations (\ref{plasma_cont_long_scale}) and (\ref{par_mom_long_scale}), derived in the regime ${\rho_i'}^2\to 0$, do not provide an accurate description at ion scales, i.e. when $d_i'\lesssim 1$ or $>1$. Below we construct a localized, stationary, 2-d (in cylindrical geometry) solution of the full system of model equations (\ref{charge_cont_DL})-(\ref{pres_bal_DL}), assuming that its spatial extent is comparable with ion scales (i.e. with the ion collisionless skin depth, viz. $d_i'\sim 1$, and with the ion Larmor radius, viz. $\rho_i'\lesssim 1$), but much larger than the electron skin depth, $d_e'\to 0$.  
Same as in the preceding subsection, we seek a solution that is travelling with the (non scaled) velocity $\vec{u}_\bot = \vec{e}_y \, k/\omega$ and is tilted to the $z$-axis by a small angle $\theta = \omega/k u_z$, that depends only on the dimensionless variables $x'$ and $y' + (c_A/u_z)\, z' - t'$. Then, the  equations of parallel electron momentum (\ref{par_mom_elec_DL}), electron continuity (\ref{elec_cont_DL}), parallel fluid momentum (\ref{par_mom_ion_DL}), charge continuity (\ref{charge_cont_DL}), and pressure balance (\ref{pres_bal_DL}) can be conveniently cast in the following form:
\begin{eqnarray}
&&
\label{par_mom_elec_WP}\hspace{-1cm}
\left[\vec{e}_z\times\nabla_\bot\left(\psi-x\right)\right]\cdot\nabla_\bot\left[\phi - p_{e_\Vert} - x + \left(1/2\right)\left(\beta_{e_\Vert} - \beta_{e \bot}\right) \; d_i^2 B_z\right] = 0,
\\ \nonumber
\\
&&\nonumber\hspace{-1cm}
\left[\vec{e}_z\times\nabla_\bot\left(\phi-p_{e_\bot}-x\right)\right]\cdot\nabla_\bot\left(n-B_z\right)-
\\
&&
\label{elec_cont__WP}\hspace{-1cm}\left[\vec{e}_z\times\nabla_\bot\left(\psi-x\right)\right]\cdot\nabla_\bot
\left\{{V_\Vert} + \left(c_A^2/u_z^2\right)\left[1 - \left(1/2\right)\left(\beta_{e_\Vert} - \beta_{e_\bot}\right)\right] \nabla_\bot^2 \psi \right\}= 0,
\\ \nonumber
\\
&&\nonumber\hspace{-1cm}
\left[\vec{e}_z\times\nabla_\bot\left(\Phi-p_{i_\bot}+
\rho_i^2 B_z - x \right)\right] \cdot\nabla_\bot\, d_i^2 {V_\Vert} -
\\
&& \nonumber\hspace{-1cm}
\left({c_A^2/u_z^2}\right)\left[\vec{e}_z\times\nabla_\bot\left(\psi-x\right)\right]\cdot\nabla_\bot
\left[p_{e_\Vert}+p_{i_\Vert} - \left({\beta_{e_\Vert}-\beta_{e_\bot}} + \beta_{i_\Vert}-\beta_{i_\bot}\right) {\left(\rho_i^2/\beta_{i_\bot}\right)} B_z - \left({2\beta_{i_\Vert}}/{\beta_{i_\bot}}-1\right)\rho_i^2\nabla_\bot^2\Phi\right] =
\\
&& \label{par_fluid__WP}
-\rho_i^2\left({c_A^2/u_z^2}\right)\left(\vec{e}_z\times\nabla_\bot \; {\partial\Phi}/{\partial x_i}\right)\cdot\nabla_\bot{\partial \psi}/{\partial x_i},
\\ \nonumber
\\
&&\nonumber\hspace{-1cm}
\left[\vec{e}_z\times\nabla_\bot\left(-\rho_i^2\nabla_\bot^2\Phi/2 + p_{i_\bot} + p_{e_\bot}\right)\right]\cdot\nabla_\bot\left(n-B_z\right) +
\\
&&\nonumber\hspace{-1cm}
\left(c_A^2/u_z^2\right)\left[1 - \left(1/2\right)\left(\beta_{e_\Vert} - \beta_{e_\bot} + \beta_{i_\Vert} - \beta_{i\bot}\right)\right]\left[\vec{e}_z\times\nabla_\bot\left(\psi-x\right)\right]\cdot\nabla_\bot
\nabla_\bot^2 \psi
{+ \rho_i^2 \left(\vec{e}_z\times\nabla_\bot V_\Vert\right)\cdot\nabla_\bot\nabla_\bot^2 \psi}
-
\\ \nonumber
&&
\hspace{-1cm}
\left[\vec{e}_z\times\nabla_\bot\left(\Phi-p_{i_\bot}+
\rho_i^2 B_z - x \right)\right] \cdot\nabla_\bot\nabla_\bot^2\Phi
+
{\rho_i^2 \left[\vec{e}_z\times\nabla_\bot\left(\psi-x\right)\right]\cdot\nabla_\bot\nabla_\bot^2 {V_\Vert}} =
\\
&&\label{charge_cont_WP}
\left[\vec{e}_z\times\nabla_\bot\left(\partial/\partial x_i\right)\left(\rho_i^2 B_z - p_{i_\bot}\right)\right] \cdot\nabla_\bot\,\partial \Phi/\partial x_i ,
%
%
\\ \nonumber
\\
&&\hspace{-1cm}
\label{pres_bal_WP}
d_i^2 B_z + p_{i_\bot} + p_{e_\bot} - \rho_i^2\nabla_\bot^2\Phi/2 = 0,
\end{eqnarray}
where, as before, we have omitted the primes and have used the following notations
\begin{equation}\label{definicije}
\Phi = \left(1 + \rho_i^2\nabla_\bot^2/2\right)^{-1}\left(\phi + p_{i_\bot}\right),
\quad
\psi = \left(u_z/c_A\right) \,  A_z
{
\quad {\rm and} \quad
V_\Vert = \left(c_A/u_z\right) \, U_{i_\Vert}
}.
\end{equation}
Electron and ion pressures are related with the density by the equation of state (\ref{eq_state_DL}), $p_{j\,\zeta} = \gamma_{j\, \zeta} \, (\beta_{j\, \zeta}/2) \; d_i^2 n$, where $j = e, i$ and $\zeta = \Vert, \bot$.
It is worth noting that the linearized version of our basic equations (\ref{par_mom_elec_WP})-(\ref{pres_bal_WP})
in the regime of small but finite ion Larmor radius corrections, $\rho_i^2\nabla_\bot^2\sim\sqrt{\epsilon}< 1$, $\rho_i^4\nabla_\bot^4 \sim\epsilon\to 0$, {and in the absence of acoustic perturbations that occurs when $u_z> {\rm max}\,(v_{Ti_\bot},\; \sqrt{\gamma_{Ti_\Vert}}\, v_{Ti_\Vert})$,} reduces to
\begin{equation}\label{linear_disp_rel}
\left(\nabla_\bot^2-\kappa^2\right)\nabla_\bot^2\psi=0,
\end{equation}
where
\begin{eqnarray}
\nonumber
&& \hspace{-1cm}
\kappa^2 = \frac{2}{\rho_i^2}\left(1-\frac{2 \, u_z^2/c_A^2}{2-\beta_{e_\Vert}+\beta_{e_\bot}-\beta_{i_\Vert}+\beta_{i_\bot}}\right) \times
\\
&&  \hspace{-1cm}
\left\{
1 - \frac{\beta_{e_\Vert}-\beta_{e_\bot}}{2} +
\left[
1 + \frac{\left(2-\beta_{e_\Vert}+\beta_{e_\bot}\right) \left(4/\beta_{i_\bot}\right)}
{2 - \beta_{e_\Vert} + \beta_{e_\bot} - \beta_{i_\Vert} + \beta_{i_\bot}}\right]
\left[
\frac{\beta_{e_\bot}+\beta_{e_\Vert}\left(\gamma_{e_\Vert}-1\right) + \beta_{i_\bot}\gamma_{i_\bot}}{2 + \beta_{e_\bot}\gamma_{e_\bot} + \beta_{i_\bot}\gamma_{i_\bot}} +
\frac{\beta_{e_\Vert}-\beta_{e_\bot}}{2}
\right]
\right\}^{-1}.
\label{kappa_complete}
\end{eqnarray}
The linear response has the form of waves when $\kappa^2<0$. The response consists of two modes, whose wavenumbers are equal to zero and to $i\kappa$. These are identified as the large-$\beta$ versions of the shear Alfv\'{e}n wave [actually, the latter features a finite perpendicular wavenumber on the electron scale, for a simple example in the case of a very-low-$\beta$ plasma  see \citep{1987JPlPh..37...81J,1994PhPl....1.2614J})] and of the kinetic slow mode, respectively. We expect that in the nonlinear regime there may exist two nonlinear vortex modes analogous to these.

A simple analysis shows that the last term in Eq. (\ref{kappa_complete}) is {strictly positive when the ion temperature is anisotropic with} $\beta_{i_\Vert}\leq \beta_{i_\bot}$, if the parallel electron temperature is not exceptionally high, $\beta_{e_\Vert}\leq 2+\beta_{e_\bot}$. This infers that the kinetic slow wave is evanescent ($\kappa^2>0$) only if its parallel phase velocity is sufficiently small,
\begin{equation}\label{evanescent}
u_z^2/c_A^2 < 1-\left(\beta_{e_\Vert}-\beta_{e_\bot}+\beta_{i_\Vert}-\beta_{i_\bot}\right)/2 \;\, {\sim {\cal O}\left(1\right)},
\end{equation}
while for larger $u_z$ we have a propagating wave.
Temperature anisotropies are common in space plasmas, see Appendices \ref{solwin} and \ref{magsh} for examples in the solar wind and the Earth's magnetosheath.

Seeking travelling/tilted solutions it is possible to integrate also the parallel electron momentum equation (\ref{par_mom_elec_WP}), viz.
\begin{equation}\label{par_mom_elec_WP_res}
\phi - x =  p_{e_\Vert} - \left(1/2\right)\left(\beta_{e_\Vert} - \beta_{e \bot}\right) \; d_i^2 B_z + F\left(\psi-x\right),
\end{equation}
where $F$ is an arbitrary function of the nonlinear characteristic $\psi-x$.

It is difficult to proceed further because solving Eq. (\ref{elec_cont__WP}) in a general case is a formidable task. Likewise, Eqs. (\ref{par_fluid__WP}) and (\ref{charge_cont_WP}) contain higher derivatives of unknown functions on their right hand sides and are practically impossible to integrate. In particular, the coupling with acoustic perturbation increases the complexity of our equations and makes an analytic solution virtually impossible. For this reason, we will restrict our study to the {perturbations whose parallel phase velocity satisfies the conditions $(c_a^2/u_z^2)(\beta_{i\bot}/2) \lesssim (c_a^2/u_z^2)(\gamma_{i\Vert}\beta_{i\Vert}/2) \sim \epsilon\to 0$ or equivalently $u_z> {\rm max}\, (v_{Ti_\bot},\; \sqrt{\gamma_{i_\Vert}}\; v_{Ti_\Vert})$ when, according to Eq. (\ref{par_fluid__WP}), we can neglect the parallel fluid velocity and set $V_\Vert\to 0$.}

\subsubsection{Large-$\beta$ shear Alfv\'{e}n solution with  ${\delta n}/{n_0} = \delta B_z/B_0$ and $E_\Vert = 0$} \label{shearAlfen}

First, we exclude acoustic waves from our analysis adopting $u_z> {\rm max}\,(v_{Ti_\bot},\, \sqrt{\gamma_{i_\Vert}}\, v_{Ti_\Vert})\; \Rightarrow\; V_\Vert\to 0$, cf. Eq. (\ref{par_fluid__WP}).
In such regime we have $\gamma_{i_\bot} = 2$ and $\gamma_{e_\bot}=1$, see Eq. (\ref{pressure}) and the subsequent discussion. An analytic solution can be constructed  following the standard Larichev{\&}Reznik procedure only in the special case when the relative perturbations of density and compressional magnetic field are equal (i.e. $n=B_z$ in dimensionless units) when, {after making use of Eq. (\ref{pres_bal_WP}),} both continuity equations (\ref{elec_cont__WP}) and (\ref{charge_cont_WP}) are drastically simplified to 2-D Euler equations, viz.
\begin{equation}\label{integr_cont_eqs}
\left[\vec{e}_z\times\nabla_\bot\left(\psi-x\right)\right]\cdot\nabla_\bot\nabla_\bot^2 \psi = 0,
\quad {\rm and} \quad
\left[\vec{e}_z\times\nabla_\bot\left(\Phi-x\right)\right]\cdot\nabla_\bot\nabla_\bot^2 \Phi = 0,
\end{equation}
that can be readily integrated as
\begin{equation}\label{Chaply1}
\nabla_\bot^2\psi = G\left(\psi-x\right),
\quad\quad
\nabla_\bot^2\Phi = H\left(\Phi-x\right).
\end{equation}
Same as before, $G$ and $H$ are arbitrary function of their argument, which we take to be part-by-part linear. Thus,  $G(\xi)$ is adopted as 
$
G(\xi)=(\xi - \xi_0)\, G_1,
$ 
with the constants $\xi_0$ and $G_1$ taking different values $\xi_0^{in}, G_1^{in}$ and $\xi_0^{out}, G_1^{out}$ inside and outside, respectively, of a (moving) vortex core defined by $x^2+ [y+ (c_A/u_z)\, z - t]^2=r_0^2$. Obviously, for a spatially localized solution we must have $G_1^{out} = 0$, while $G_1^{in}$ will be determined from the smoothness of the vector potential $\psi$ (i.e. from the absence of the $z$-component of the surface current) at the edge of the vortex core. The function $G(\psi-x)$, and consequently the parallel current $-\nabla_\bot^2\psi$, can have a finite jump for some value of $\psi-x$, see discussion following Eq. (\ref{part-by-part}) in the preceding subsection.
Now we can readily write the solution of the 2-d Euler equation (\ref{Chaply1}) as a Chaplygin vortex, identical to that obtained in the preceding subsection, Eq. (\ref{celo_resenje_ion_scale}), viz.
\begin{equation}
\label{celo_resenje_ion_scale_2}
\psi\left(r, \varphi\right) =
\left\{\begin{array}{ll}
   \cos\varphi \; \left(r_0^2/r\right), & \hbox{$r\geq r_0$,} \\
   \cos\varphi \; \left\{r - (2 r_0/j_1) \; [J_1\left(j_1 \; r/r_0\right)/J_1'\left(j_1\right)]\right\}
   + \psi_0\left[J_0\left(j_1 \; r/r_0\right) - J_0\left(j_1\right)\right] ,  & \hbox{$r<r_0$,}
\end{array} \right.
\end{equation}
The density $n$ and the compressional magnetic field $B_z$ are calculated from the generalized pressure balance (\ref{pres_bal_WP}), viz.
\begin{equation}\label{gustina}
n = B_z = \left(1 +
{\beta_{i_\bot}+\frac{\beta_{e_\bot}}{2}
}
\right)^{-1} \frac{\beta_{i_\bot}}{4} \; \nabla_\bot^2 \Phi
,
\end{equation}
which after substitution into the parallel electron momentum equation (\ref{par_mom_elec_WP}) and using (\ref{Chaply1}) 
yields
$\Phi-x = F\left(\psi-x\right)$
where $\psi$ is given by Eq. (\ref{celo_resenje_ion_scale_2}).
%
Obviously, for $r>r_0$ the slope of the function $F$ is given by $F_1=1$, which yields $n=0$ outside of the vortex core. The definition of the streamfunction $\Phi$ further implies that outside of the vortex core we also have $\phi=\psi$. As these potentials satisfy the same continuity conditions at the core edge, we must have $\phi=\psi$ on the entire $x,y$ plane, which corresponds to $E_\Vert=0$, {i.e. $\phi= u_z A_z$  in dimensional units}.
One easily sees that the dipolar components of the potentials $\phi$ and $\psi$ (that are $\propto \cos\varphi$) are continuous and smooth functions and that the corresponding density $n$ and compressional magnetic field $B_z$ are continuous at $r=r_0$. Conversely, the monopolar 
component of the compressional magnetic field has a finite jump $\Delta B_z^{(0)}$ at the edge of the vortex core, where
\begin{equation}\label{jump_B_0}
\Delta B_z^{(0)}= \psi_0\;\beta_{i_\bot}\left({j_1^2}/{4 r_0^2}\right)
\left[1 +
{\beta_{i_\bot}+\beta_{e_\bot}/2}
 + \left(1 + 
{\beta_{e_\bot}}/{2}\right) \; \left({\rho_i^2 j_1^2}/{2 r_0^2}\right)\right]^{-1}.
\end{equation}
Such discontinuity corresponds to a surface current (with zero thickness and infinite density) located at $r=r_0$, flowing in the poloidal direction. The latter is regarded as nonphysical and it probably gives rise to an instability of the Chaplygin's monopolar component. In other words, the monopolar Chaplygin component may exist only when the compressional magnetic field is negligible, that is usually the case when the plasma $\beta$ is small.
The ion-scale shear Alfv\'en vortex (\ref{celo_resenje_ion_scale_2})-(\ref{gustina}) is very similar to the {RMHD-scale}
shear Alfv\'en vortex found in the preceding Subsection and they are both displayed in Fig. \ref{slika1}. Both have a vanishing parallel electric field, corresponding to $\phi=(u_z/c_A)\, A_z$ and outside of the vortex core they have $\nabla_\bot^2\phi =\nabla_\bot^2 A_z = n = B_z = 0$. Inside the core, large-scale structures feature $n/B_z = {\rm constant}\,\ne 1$, while on the ion-scale we have been able to find analytically only vortices with $n/B_z = 1$.

In the regime when both the parallel electric field and the ion sound are absent, $E_\Vert = 0$ and {$u_z> v_{Ti_\bot},\; \sqrt{\gamma_{i_\Vert}}\; v_{Ti_\Vert}$}, the ions are thermalized in the perpendicular direction, $d/dt\ll v_{T i_\Vert} \, |\nabla_\bot|$, and the density and the parallel magnetic field perturbations are fully correlated, ${\delta n}/{n_0} = \delta B_z/B_0$, the energy conservation can be readily obtained from the charge continuity equation for arbitrary dependence on $t$ and $z$ that has the simple form [cf. Eq. (\ref{integr_cont_eqs})]:
\begin{equation}\label{ch_con_shAlf}
\left[{\partial}/{\partial t} + \left(\vec{e}_z\times\nabla_\bot\Phi\right)\cdot\nabla_\bot\right]\nabla_\bot^2 \Phi = 0.
\end{equation}
Multiplying by $\Phi$ and integrating for the entire space (provided $\Phi= 0$ at the infinity) we obtain $\partial W/\partial t =0$, where
\begin{equation}\label{energy_shear_alf}
W = \left({1}/{2}\right)\int_{-\infty}^\infty\int_{-\infty}^\infty\int_{-\infty}^\infty dx \, dy\, dz \left(\nabla_\bot{\Phi}\right)^2 .
\end{equation}

\subsubsection{Large-$\beta$ kinetic slow magnetosonic solution, with $\delta n/n_0 \ne \delta B_z/B_0$, $u_z< c_A$, and $\vec{B}_\bot=0$} \label{kineticAlfen}

{In the regime with negligible contribution of acoustic perturbation defined above, when $V_\Vert\to 0$, $\gamma_{i_\bot}= 2$, $\gamma_{e_\bot}=1$,
and  $u_z>  {\rm max}\, (v_{T i_\bot}, \sqrt{\gamma_{i_\Vert}} v_{T i_\Vert})$}, we seek a travelling solution whose perturbations of density and compressional magnetic field are not fully correlated, i.e. with $n\ne B_z$. As outside of the vortex core the localized nonlinear solution is essentially a linear evanescent response to the nonlinearities located within the core, we ascertain from Eq. (\ref{kappa_complete}) that in the plasma regimes of interest, featuring ion temperature anisotropy, $\beta_{i_\Vert}\leq \beta_{i_\bot}$ and a moderate parallel electron temperature, $\beta_{e_\Vert}\leq 2+\beta_{e_\bot}$, the parallel phase velocity of kinetic slow mode vortices can not be much bigger than the Alfv\'{e}n speed, see Eq. (\ref{evanescent}). We adopt a somewhat more rigorous restriction for $u_z$, viz. $c_A > u_z > {\rm max}\, (v_{T i_\bot}, \sqrt{\gamma_{i_\Vert}} v_{T i_\Vert})$ that permits us to simultaneously set ${V_\Vert}\ll 1$ and $\psi\sim u_z^2/c_A^2\, \ll 1$. The corresponding kinetic slow wave is localized, $\kappa^2>0$ see Eq. (\ref{evanescent}), and it can be realized when the perpendicular ion temperature is of the order $\beta_{i_\bot}\lesssim 1$ and the parallel ion temperature is sufficiently small, $v_{T i_\Vert}^2/c_A^2 = \beta_{i_\Vert}/2 \ll 1$. Such ordering is easy to achieve in the Earth's magnetosheath downstream of a quasi-perpendicular bow shock, {possibly also in the fast solar wind, but more difficult in the slow} solar wind where the separation between $c_A$ and $v_{Ti_\Vert}$ is smaller, see Appendices \ref{solwin} and \ref{magsh}.
Besides, if the parallel electron temperature is not extremely small, $T_{e_\Vert}/T_{i_\Vert}\geq m_e/m_i$, the electrons are isothermal along the magnetic field, too.
In such regime we have $\gamma_{e_\Vert} = \gamma_{e_\bot} = 1$, and {making use of Eq. (\ref{par_mom_elec_WP_res})} we can rewrite the electron continuity equation (\ref{elec_cont__WP}) as
\begin{eqnarray}
\nonumber
&&
\left[\vec{e}_z\times\nabla F\left(\psi-x\right)\right]\cdot\nabla\left(n-B_z\right) -
\frac{c_A^2}{u_z^2}\left(1 - \frac{\beta_{e_\Vert} - \beta_{e_\bot}}{2}\right) \left[\vec{e}_z\times\nabla\left(\psi-x\right)\right]\cdot\nabla \nabla_\bot^2\psi =
\\
&&
\label{nova_druga}-
\left[\vec{e}_z\times\nabla\left(p_{e_\Vert}-p_{e_\bot} - \frac{\beta_{e_\Vert}-\beta_{e_\bot}}{2}\; d_i^2 B_z\right)\right]\cdot\nabla\left(n-B_z\right).
\end{eqnarray}
We readily note that the right hand side of this equation vanishes for isothermal electrons. Namely, by virtue of the isothermal equation of state (\ref{eq_state_DL}), we have $p_{e_\Vert}-p_{e_\bot}=(1/2)(\beta_{e_\Vert}-\beta_{e_\bot})\,d_i^2 \, n$ and the right-hand-side of Eq. (\ref{nova_druga}) reduces to zero as a mixed product of two colinear vectors. This enables the equation to be integrated as
\begin{equation}\label{elec_cont__WP_integr}
n - B_z = \frac{c_A^2}{u_z^2}\left(1 - \frac{\beta_{e_\Vert} - \beta_{e_\bot}}{2}\right)\frac{\nabla_\bot^2\psi}{F'\left(\psi - x\right)} + H\left(\psi - x\right),
\end{equation}
where $H$ is an arbitrary function. It can be shown that solutions with arbitrary $u_z/c_A$ can meet all physical continuity conditions at the core's edge only if they contain both the shear- and the kinetic slow mode components described in Eq. (\ref{kappa_complete}). However, in such a case the nonlinear term on the right hand side Eq. (\ref{charge_cont_WP}) is finite on the entire $x-y$ plane, which presents a formidable obstacle for an analytic treatment and requires extensive numerical calculations that are outside the scope of the present paper. To proceed, we restrict ourselves to the phase velocities that are much smaller than the Alfv\'{e}n speed, $c_A\gg u_z > v_{Ti_\Vert} $, when a solution involving only one Alfv\'{e}n mode becomes possible (in the solar wind event recorded by Cluster \citep{Denises_paper} such strong constraint might not be fulfilled, cf. Appendix \ref{solwin}, and those structures are likely to be coupled either with the ion sound or with the torsional magnetic field). In such regime, the electron continuity (\ref{elec_cont__WP}) gives $\psi\sim (u_z^2/c_A^2) \; \phi \to 0$, which in turns yields that the arguments of the functions $F$ and $G$ reduces to $\xi \equiv \psi-x \to -x$. Same as before, these functions are adopted to be part-by-part linear, in the form Eq. (\ref{part-by-part}), where the slopes $F_1$ and $H_1$ take different constant values $F_1^{in},\, H_1^{in}$,  and $F_1^{out}, H_1^{out}$ inside and outside of the vortex core determined by $\xi(r_0) = \xi_0$. We note the separatrix $r=r_0$ is not an isoline of the functions $F$ and $H$, whose argument is given by $\xi = \psi-x\to -x = -r\,\cos\varphi$, which obviously is not constant at the separatrix $r=r_0$. This prohibits the slopes to jump at the circle $r=r_0$ and implies that $F_1^{in} = F_1^{out} = 1$ and $H_1^{in} = H_1^{out} = 0$. As a consequence, Eq. (\ref{elec_cont__WP_integr}) is decoupled from the rest of the system, while from Eq. (\ref{par_mom_elec_WP}) we readily obtain
\begin{equation}\label{par_mom_elec_WP_psi_0}
\phi - p_{e_\Vert} + \left(1/2\right)\left(\beta_{e_\Vert} - \beta_{e \bot}\right) \; d_i^2 B_z = 0.
\end{equation}
Setting $\gamma_{i_\bot}=2$ and $\gamma_{e_\Vert}$, the quantities $n$, $B_z$, and $\phi$ can now be expressed from Eqs. (\ref{pres_bal_WP}),  (\ref{definicije}), and (\ref{par_mom_elec_WP_res}) as follows
(for easier reading, here and below we use the \texttt{mathcal} font to denote true constants, such as  ${\cal N}, {\cal B},{\cal F},{\cal Q},{\cal A}$, and ${\cal U}$, that depend only on the plasma parameters and \textit{not} on the slopes $G_1^{in}$ and $G_1^{out}$)
\begin{eqnarray}
\label{n_drift}
&&n = {\cal N}_0\Phi + {\cal N}_2 \nabla_\bot^2\Phi \equiv
\frac{1}{\rho_i^2{\cal Q}}\;\Phi \, +\, \frac{2+\beta_{e_\Vert}-\beta_{e_\bot}}{4{\cal Q}} \; \nabla_\bot^2 \Phi,
\\
\label{B_drift}
&&B_z = {\cal B}_0\Phi + {\cal B}_2 \nabla_\bot^2\Phi \equiv
-\frac{{2}\beta_{i_\bot} + \beta_{e_\bot}}{2\rho_i^2{\cal Q}}\;\Phi\, +
\frac{\beta_{e_\Vert}-\beta_{e_\bot}}{4{\cal Q}}\;\nabla_\bot^2\Phi,
\\
\label{phi_drift}
&&\phi = {\cal F}_0\Phi + {\cal F}_2 \nabla_\bot^2\Phi \equiv
\frac{{\cal Q}-{2}}{\cal Q}\; \Phi \, + \, \frac{{\cal Q}-{2 - 
\beta_{e_\Vert}
+ \beta_{e_\bot}
}}{2{\cal Q}}\;\rho_i^2\; \nabla_\bot^2\Phi,
\\
&&{\cal Q} = \frac{
\beta_{e_\Vert} + {2}\beta_{i_\bot}}{\beta_{i_\bot}} +
\frac{{2}\beta_{i_\bot}+\beta_{e_\bot}}{2} \; \frac{\beta_{e_\Vert} - \beta_{e_\bot}}{\beta_{i_\bot}}
\quad\quad {\rm and} \quad\quad
{\cal U}= 
1-\frac{\beta_{i_\Vert}-\beta_{i\bot}}{2 - \beta_{e_\Vert} + \beta_{e_\bot}}
\label{def_D_U}
\end{eqnarray}
It is worth noting that due to the ions' FLR effects, Eq. (\ref{B_drift}) implies that the stream function $\Phi$ is not proportional to the magnetic field $B_z$, which essentially decouples the velocity and magnetic fields.
Using Eqs. (\ref{par_mom_elec_WP_res}) and (\ref{elec_cont__WP_integr}) and some simple manipulations, the above expressions permit us to rewrite the charge continuity equation (\ref{charge_cont_WP}) as follows
\begin{eqnarray}
&&
\nonumber
\left[\vec{e}_z\times\nabla\left(n-B_z\right)\right]\cdot\nabla\left[\Phi - {\cal U}\; x
-{\cal A}_1 \,\rho_i^2\left(n-B_z\right)
\right] +
\left\{\vec{e}_z\times\nabla\left[\Phi-x-\rho_i^2 \left(n - B_z\right)
+{\cal A}_2 \, \rho_i^2\,\nabla_\bot^2 \Phi
\right]\right\}\cdot \nabla\nabla_\bot^2 \Phi = 
\\
&&
\label{charge_cont_WP_2}
\left[\vec{e}_z\times\nabla_\bot\left(\partial/\partial x_i\right)\rho_i^2\left(n-B_z\right)\right] \cdot\nabla\left(\partial \Phi/\partial x_i\right),
\end{eqnarray}
%
where ${\cal A}_1$ and ${\cal A}_2$ are arbitrary constants introduced for algebraic convenience.
Adopting these in the following way
\begin{equation}\label{alfa1_alfa2}
{\cal A}_1 = {\cal U}-\frac{{\cal U} {-{\cal U}/{\cal Q}} -1}{\rho_i^2\left({\cal N}_0-{\cal B}_0\right)},
\quad\quad
{\cal A}_2 = \frac{{\cal U} {-{\cal U}/{\cal Q}} -1}{\cal U} \,\frac{{\cal N}_2-{\cal B}_2}{\rho_i^2\left({\cal N}_0-{\cal B}_0\right)},
\end{equation}
using Eqs. (\ref{n_drift}) and (\ref{B_drift}), and after some algebra, we can cast Eq. (\ref{charge_cont_WP_2}) in a simple form, viz.
\begin{equation}
\label{charge_cont_WP_3_integr}
\left[\vec{e}_z \times\nabla\left(\Phi + {\cal V}\,x\right)\right]\cdot\nabla\left(\nabla_\bot^2\Phi + \kappa^2{\cal V}\,x\right) =
2\,{\cal C} \left(\vec{e}_z \times\nabla\;\partial\Phi/\partial x_i\right)\cdot \nabla\nabla_\bot^2\partial\Phi/\partial x_i ,
\end{equation}
where
\begin{equation}\label{kappa_v_C}
\kappa^2 = \frac{{\cal U}\left({\cal N}_0 - {\cal B}_0\right)}{1-{\cal U}\left({\cal N}_2 - {\cal B}_2\right)},
\quad\quad
{\cal V} = -\frac{1-{\cal U}\left({\cal N}_2-{\cal B}_2\right)}{1 -{ 1/{\cal Q}} - \rho_i^2 \left({\cal N}_0 -{\cal B}_0\right)- \left({\cal N}_2- {\cal B}_2\right)}
\quad\quad {\rm and} \quad\quad
{\cal C} =\frac{\rho_i^2\kappa^2}{2}\;\frac{\cal V}{\cal U}\;\;\frac{{2}{\cal N}_2-{\cal B}_2}{{\cal N}_0-{\cal B}_0}.
\end{equation}
In the regime of small but finite FLR corrections ${\cal C}\nabla_\bot^2\ll 1$, the right-hand-side of Eq. (\ref{charge_cont_WP_3_integr}) comprises a small correction and, iteratively, it can be approximated by using the leading order solution of Eq. (\ref{charge_cont_WP_3_integr})
$
\nabla_\bot^2\Phi \approx -\kappa^2{\cal V}\,x + G(\Phi + {\cal V}\,x)
$, where $G$ is an arbitrary function of its argument. Then, using the identity
\begin{equation}\label{identity}
2 \, \nabla_\bot\cdot\left\{\left[\left(\vec{e}_z\times\nabla_\bot f\right)\cdot\nabla_\bot\right]\nabla_\bot G\left(f\right)\right\} = \left(\vec{e}_z\times\nabla_\bot f\right)\cdot\nabla_\bot\nabla_\bot^2 G\left(f\right) - \left[\vec{e}_z\times\nabla_\bot\nabla_\bot^2 f\right]\cdot\nabla_\bot G\left(f\right),
\end{equation}
we can rewrite Eq. (\ref{charge_cont_WP_3_integr}) in the following form
\begin{equation}
\label{charge_cont_WP_4_integr}
\left[\vec{e}_z \times\nabla\left(\Phi + {\cal V}\,x\right)\right]\cdot\nabla\left[\nabla_\bot^2\Phi + \kappa^2{\cal V}\,x + {\cal C}\nabla_\bot^2 G\left(\Phi + {\cal V}\,x\right)\right] =
\left[\vec{e}_z \times\nabla G\left(\Phi + {\cal V}\,x\right)\right]\cdot\nabla\;{\cal C}\nabla_\bot^2\left(\Phi + {\cal V}\,x\right),
\end{equation}
that is, with the accuracy to the first order in the small quantity ${\cal C}\nabla_\bot^2$, equivalent to
\begin{equation}
\label{charge_cont_WP_4_equiv}
\left[\vec{e}_z \times\nabla\left(1+{\cal C}\nabla_\bot^2\right)\left(\Phi + {\cal V}\,x\right)\right]\cdot\nabla\left[\left(1+{\cal C}\nabla_\bot^2\right) \left(\nabla_\bot^2\Phi + \kappa^2{\cal V}\,x\right) \right] =0,
\end{equation}
and is readily integrated one time, viz.
\begin{equation}\label{charge_cont_WP_5_integr_2}
\left(1+{\cal C}\;\nabla_\bot^2\right) \left(\nabla_\bot^2\Phi + \kappa^2{\cal V}\,x\right) = G\left[\left(1+{\cal C}\;\nabla_\bot^2\right) \left(\Phi + {\cal V}\,x\right)\right].
\end{equation}
We adopt $G(\xi)$ in the form of a continuous part-by-part linear function, $G(\xi)=G_1 \xi$, whose constant slope $G_1$ takes different values $G_1^{in}$ and $G_1^{out}$ inside and outside of the circle $r=r_0$, respectively. Remarkably, with such choice of $G(\xi)$, the parameter $\cal C$ coming from the nonlinear term on the right-hand-side of Eq. (\ref{charge_cont_WP_3_integr}) and (\ref{charge_cont_WP}), cancels out in Eq. (\ref{charge_cont_WP_5_integr_2}) and does not affect its solution. As the function $G(\xi)$ must be continuous [see discussion in the paragraph following Eq. (\ref{nonlin_disp_rel}) at the end of this section], a jump is permitted only if the argument vanishes at such circle, $\xi(r_0,\varphi)=0$. Noting that for 
a localized solution we must have $G_1^{out}=\kappa^2$, and setting $G_1^{in}=-\lambda^2$, we obtain the following equations for the stream function $\Phi$ outside and inside the circle $r=r_0$,
\begin{eqnarray}
\nonumber
&&
\left(\nabla_\bot^2-\kappa^2\right)\Phi^{out} = 0, 
\hspace{2cm} r>r_0,
\\
&&
\left(\nabla_\bot^2 + \lambda^2\right)
\left[\Phi^{in}+\left(1+\frac{\kappa^2}{\lambda^2}\right){\cal V} \, x\right] = 0, 
\hspace{.3cm}
r<r_0.
\label{jednacina_napolju_unutra}
\end{eqnarray}
These separate variables in cylindrical coordinates, $\Phi=\sum_k\Phi_k\exp(ik\varphi)$; amplitude of the $k$-th harmonic is given by
\begin{equation}
\Phi^{out}_k =c_k^{out}\, K_k\left(\kappa r\right),
\quad\quad
\Phi^{in}_k = c_k^{in} \, J_k\left(\lambda r\right),
\end{equation}
where $c_k^{in}$ and $c_k^{out}$ are arbitrary constants.
It can be argued that the stream function $\Phi$ must be a dipole, i.e. that it may contain only the dipole component $k=1$ (for a discussion, see the paragraph at the end of this Section). Then, the continuity of the function $G$ readily yields
\begin{equation}\label{continuity_Phi}
\left(\Phi^{out}+{\cal V}\,x\right)_{r=r_0} = \left(\Phi^{in}+{\cal V}\,x\right)_{r=r_0} =
\left(\nabla_\bot^2\Phi^{out}+\kappa^2{\cal V}\,x\right)_{r=r_0} =
\left(\nabla_\bot^2\Phi^{in}+\kappa^2{\cal V}\,x\right)_{r=r_0} = 0,
\end{equation}
which from Eqs. (\ref{n_drift})-(\ref{phi_drift}) provides also the continuity of the functions $n$, $B_z$, and $\phi$. Finally, matching the above "in" and "out" solutions at $r=r_0$ we obtain a standard Larichev\& Reznik-type dipole \citep{Larichev_Reznik}
\begin{equation}
\label{solution_in_out}
\Phi\left(r, \varphi\right) = {\cal V}\, r_0\;\cos\varphi \times
\displaystyle{
\left\{\begin{array}{ll}
  -\frac{K_1\left(\kappa r\right)}{K_1\left(\kappa r_0\right)}, & \hbox{$r>r_0$} \\
  -\left(1+\frac{\kappa^2}{\lambda^2}\right)\frac{r}{r_0} + \frac{\kappa^2}{\lambda^2}\frac{J_1\left(\lambda r\right)}{J_1\left(\lambda r_0\right)},  & \hbox{$r<r_0$}
\end{array} \right. },
\end{equation}
while the plasma density, compressional magnetic field and the electrostatic potential are expressed from Eqs. (\ref{n_drift})-(\ref{phi_drift}).
%
This solution is localized in space only if the '$out$' e-folding length $\kappa$ defined in Eq. (\ref{kappa_v_C}) or, equivalently obtained from Eq. (\ref{kappa_complete}) in the limit $u_z\ll c_A$ and $\gamma_{e_\Vert}=1$, is a real quantity, i.e. for $\kappa^2>0$ which yields the condition for the existence of kinetic slow mode vortices with a complicated dependence on the values of plasma $\beta_{j_\zeta}$, with $j=e,i$ and $\zeta =\Vert,\bot$. In contrast to its shear Alfv\'en counterparts, Eqs. (\ref{celo_resenje_ion_scale}) and (\ref{celo_resenje_ion_scale_2}), which are essentially MHD nonlinear modes and do not have a spatial scale, the 
kinetic slow mode vortex Eq. (\ref{solution_in_out}) has a distinct scale comparable the ion Larmor radius $1/\kappa\sim \rho_i$, {see Eqs. (\ref{kappa_complete}), (\ref{kappa_v_C}). Also, linear equations indicate that outside of the vortex core perturbations of the compressional magnetic field and density of a kinetic slow mode vortex are anticorrelated, $n/B<0$.}

The remaining free parameter $\lambda$ is determined from the condition that the radial electric field is continuous at the edge of the core, $
(\partial\phi^{in}/\partial r)_{r = r_0} = 
(\partial \phi^{out}/\partial r)_{r = r_0}$, i.e. of the absence at $r=r_0$ of any surface charges. This gives rise to the following nonlinear dispersion relation
\begin{equation}\label{nonlin_disp_rel}
\left({\cal F}_0 + {\cal F}_2\kappa^2\right) \frac{\kappa r_0\, K_1'\left(\kappa r_0\right)}{K_1\left(\kappa r_0\right)} = {\cal F}_0\left(1+\frac{\kappa^2}{\lambda^2}\right) - \left({\cal F}_0 - {\cal F}_2\lambda^2\right)\frac{\kappa^2}{\lambda^2} \frac{\lambda r_0 \, J_1'\left(\lambda r_0\right)}{J_1\left(\lambda r_0\right)}.
\end{equation}
We have shown in the subsection \ref{large_scale} 
that the Chaplygin's monopole component of the solution can exist only when the function $G(\Phi+{\cal V}x)$ features a finite jump at the edge of the vortex core, which on the spatial scale sufficiently bigger than the ion Larmor radius, considered there, produces a jump in the vorticity $\nabla_\bot^2\Phi$. However, approaching the ion scales and including FLR terms, in the kinetic slow mode branch we obtain Eq. (\ref{charge_cont_WP_5_integr_2}) that contains also the Laplacian of vorticity, $\nabla_\bot^4\Phi$. The latter obviously becomes singular when $\nabla_\bot^2\Phi$ has a jump.
Such singularity, that implies also a singularity in the charge continuity equation (\ref{charge_cont_WP}) or  (\ref{charge_cont_WP_2}), is clearly prohibited for physical reasons. This indicates that (quasi)monopolar Chaplygin structures in the form (\ref{celo_resenje_ion_scale_2}) cannot develop on the ion scale, in the kinetic slow mode. As the singularity of $\nabla_\bot^4\Phi$ arises in Eq. (\ref{charge_cont_WP_4_equiv}) due to the expansion in the powers of small-but-finite FLR inherent in the stress tensor Eq. (\ref{braginskii}), one may expect that it vanishes in a plasma description that involves all orders in FLR, such as the gyrofluids. It can be expected that a gyrofluid kinetic slow magnetosonic Chaplygin vortex features a sharp peak of charge density and/or a thin current layer with a large current density that may affect its stability. Numerical study of such structures requires extensive calculations and is beyond the scope of this paper.

In a kinetic slow mode regime without ion sound and torsional magnetic field, $d/dt \gg v_{T i_\Vert} \; \partial/\partial x_b$ and  $\vec{B}_\bot=0$ $ (\,\Rightarrow d/dt \ll  c_A  \;\partial/\partial x_b)$, 
the charge continuity equation with arbitrary $(t, \vec{r}_\bot, z)$ dependence has the form [cf. Eq. (\ref{charge_cont_WP_3_integr})]:
\begin{equation}
\label{chargcont_kin_alf}
\left[{\cal V}\, \partial/\partial t + \left(\vec{e}_z \times\nabla\Phi \right)\cdot\nabla\right]\left(\nabla_\bot^2- \kappa^2\right) \Phi =
2\,{\cal C} \left(\vec{e}_z \times\nabla\;\partial\Phi/\partial x_i\right)\cdot \nabla\nabla_\bot^2\partial\Phi/\partial x_i.
\end{equation}
where $\kappa$, ${\cal V}$, and ${\cal C}$ are given in Eq. (\ref{kappa_v_C}). Multiplying Eq. (\ref{chargcont_kin_alf}) by $\Phi$ and after some algebra, we can cast it in the form
\begin{equation}
\left[{\cal V}\,\Phi\, \frac{\partial}{\partial t} + \left(\vec{e}_z \times\nabla\frac{\Phi^2}{2} \right)\cdot\nabla\right]\left(\nabla_\bot^2- \kappa^2\right) \Phi
=
{\cal C}\left\{\left(\vec{e}_z \times\nabla\nabla_\bot^2\Phi\right)\cdot \nabla\left(\nabla_\bot\Phi\right)^2 -
\frac{\partial}{\partial x_i}\left[2\,\Phi\,\left(\vec{e}_z \times\nabla\;\nabla_\bot^2\Phi\right)\cdot \nabla\frac{\partial\Phi}{\partial x_i}\right] \right\}.
\label{chargcont_kin_alf2}
\end{equation}
The term on the right-hand-side is a divergence and vanishes in the integration for the entire space, provided the effective potential $\Phi$ vanishes at infinity. Thus we obtain the expression for the energy conservation $\partial W/\partial t = 0$, where
\begin{equation}\label{energy_kinetic_Alf}
W = \left({1}/{2}\right)\int_{-\infty}^\infty \int_{-\infty}^\infty\int_{-\infty}^\infty dx \, dy\, dz \left[\left(\nabla_\bot{\Phi}\right)^2 + \kappa^2\Phi^2\right].
\end{equation}

\section{Discussions and concluding remarks}\label{Discussions}

We have studied fluid plasma vortices in a high-$\beta$ plasma, on the spatial scale comparable to the ion inertial length and approaching the ion Larmor radius, including the effects of the compression of the magnetic field and of the finite ion Larmor radius, in the regime when the acoustic perturbations are 
small. The vortices have the form of infinitely long filaments, slightly tilted to the magnetic field. Our basic Eqs. (\ref{par_mom_elec_WP})-(\ref{pres_bal_WP}) possess also a trivial stationary solution that is fully aligned with the $z$-axis, $\partial/\partial t=\partial/\partial z =0$ and circularly symmetric $\partial/\partial\varphi =0$, i.e. strictly monopolar. However, water tank experiments \citep{2003PhFl...15.1033B}, in which perturbations evolved according to the 2-d Euler equation (\ref{integr_cont_eqs}) but involved also a small but finite viscosity of the fluid (non existent in our plasma regime), revealed that such stationary monopoles either disperse or slowly transform into dipolar or tripolar vortices, depending on the amount of shielding in the initial state. This may be related also with the jumps in the vorticity, $\nabla_\bot^2\Phi$, at the edge of a monopole. Conversely, the propagating Lamb dipole, corresponding to a shear-Alfv\'en vortex with $\psi_0=0$ in Eqs. (\ref{celo_resenje_ion_scale}) and (\ref{celo_resenje_ion_scale_2}), was remarkably stable 
and could easily survive collisions with other dipoles \citep{1989Natur.340..212V}. A propagating quasi-monopolar vortex, i.e. a Chaplygin's structure with a relatively small dipolar component, described by Eq. (\ref{celo_resenje_ion_scale}) when $\psi_0\ne 0$, is much more stable than the stationary monopoles. In an ordinary fluid, a Chaplygin's quasi monopole may propagate over a distance that is an order of magnitude bigger than its diameter, as suggested by the weak nonlinear theory and numerical simulations \citep{1998JPO....28...22S}, as well as by experiments in non-rotating water tanks with rectangular shape \citep{1999JPO....29.2741V}.
More recent water tank experiments \citep{2006NPGeo..13..641C} have demonstrated that it still has a finite lifetime, because the secondary component of such strongly asymmetric vortex pair starts to wrap around the principal monopole creating a strain, that eventually gives rise to an elliptic instability due to the parametric resonance between the oscillation of inertial waves and the ambient strain field, for details see Ref. \citep{2006NPGeo..13..641C}. It should be noted that the behavior of Chaplygin vortices in a fully 3-D geometry is still an open question, since no reliable 3-D simulations and experiments have been reported in the literature and we are unable to predict whether the dynamical 3-D turbulence of the solar wind and of the Earth's magnetosheath is dominated by stable dipolar \citep{Larichev_Reznik,1999PhPl....6..713K,1992swpa.book.....P} vortices, or by long-lived, mostly monopolar Chaplygin structures. Their 
monopolar components feature a jump of the plasma density and of the compressional magnetic field $B_z$ at the edge. In the presence of a compressional magnetic field, such jumps are associated with a current shaped as a thin hollow cylinder at $r=r_0$, which probably reduces its stability in high-$\beta$ turbulent plasmas.

In a high-$\beta$ plasma, we have found two distinct types of coherent vortices propagating in the perpendicular direction. The first is identified as a generalized shear-Alfv\'{e}n structure that possesses both the torsional and the compressional component of the magnetic field perturbation. It has a zero parallel electric field and, being homogeneous along its axis that is inclined to the ambient magnetic field, it sweeps along the $z$-axis with a 
velocity $u_z$ that is in the Alfv\'{e}n speed range; the transverse phase velocity is equal to $u_z \tan\theta$, where $\theta$ is the (small) pitch angle between the structure and the background magnetic field. While in a sufficiently incompressible plasma $\delta n/n_0\to 0, \delta B_z/B_0\to 0$ 
it has the structure of a moving Chaplygin's vortex with a monopole superimposed on a dipole, in plasmas with $\beta\sim 1$ its monopolar component is likely to be unstable and short-lived due to the emergence of a thin, {(electron scale)} current layer and/or a sharp peak of charge density at its edge. The compressible magnetic field associated with such vortex is restricted to the interior of the vortex core, while the transverse perturbation "leaks" from the core to larger distances.

The second type of propagating 
structures obtainable analytically, {possesses finite compressional magnetic field and parallel electric field, as well as the perpendicular fluid velocity and the density perturbation, but vanishing parallel ion fluid velocity and the transverse perturbations of the magnetic field.} It is identified as a nonlinear kinetic slow magnetosonic structure. Its parallel phase velocity is much smaller than the Alfv\'{e}n speed which also yields a thermalized electron distribution. The transverse fluid velocity of the kinetic slow mode vortex is better localized than that of its shear-Alfv\'{e}n counterpart, while its compressional magnetic field extends outside the core. {A kinetic slow mode structure that possesses only compressional magnetic field perturbation} exists only when $v_{T i_\Vert}^2/c_A^2 = \beta_{i_\Vert}/2 \ll 1$, otherwise it involves also the ion sound or the torsional magnetic field.

Our analytical study does not exclude the possibility of mixed shear Alfv\'en/kinetic slow magnetosonic vortices whose parallel phase velocity approaches Alfv\'en speed, $u_z\lesssim c_A$, but their construction requires extensive numerical calculations that are out of the scope of the present paper.
%
Conversely, we have demonstrated that fluid-type (quasi)monopolar Chaplygin's filaments are not likely to emerge in the kinetic slow magnetoacoustic domain. The only viable propagating kinetic magnetoacoustic monopoles, possibly in the form of cigars (i.e. of filaments with a finite length), may emerge in the situations with non-vanishing $E_\Vert$ and with the parallel phase velocities in the thermal range, not studied here. They involve particles trapped both in the electrostatic potential wells and in magnetic depressions that provide an additional nonlinearity capable to produce the spatial localization of a vortex. Coherent vortex structures in a high-$\beta$ plasma, with $\delta n/n_0 \ne \delta B_z/B_0$ and with a finite parallel electric field, $\phi\ne (u_z/c_A)\, A_z$, that include kinetic phenomena such as particle trapping has been studied elsewhere \citep{buduci_rad_sa_Deniz} via a high-$\beta$ gyrokinetic theory.
Our results can explain observations of the solar wind and the magnetosheath turbulence in a plasma with $\beta\sim 1$, in particular the Alfv\'en vortices and the compressible magnetic filaments. The structures at large scales (${\cal L}\sim 30 \rho_i$) and at the ion scales (${\cal L}\gtrsim \rho_i\sim c/\omega_{pi}$), described in Subsections \ref{large_scale} and \ref{ion_scales}, can be an important ingredient of the kinetic turbulent cascade. The latter produces power law spectra of $\delta B$-fluctuations $\sim k^{-5/3}$ and $\sim k^{-2.8}$ at MHD and ion scales, respectively \citep{
2009PhRvL.103p5003A,Denises_paper,2017ApJ...849...49P}, while at the dissipative (electron) scale the dependence is exponential $\sim\exp(-k\rho_{Le})$, \citep{2012ApJ...760..121A}.
{The compressible component of the inertial range solar wind turbulence at 1 AU has been shown \citep{2012ApJ...753L..19H} to belong almost entirely to the kinetic slow mode, which determines the nature of the density fluctuation spectrum and of the the cascade of kinetic turbulence to short wavelengths.}

\appendix

\section{Plasma parameters in the solar wind}\label{solwin}

Plasma parameters in the region of the slow solar wind where vortex structures were observed \citep{Denises_paper} are listed below.
Such plasma can be regarded as weakly magnetized, since $|\Omega_e|/\omega_{p e} \sim 0.0056$, $\Omega_i/\omega_{p i} \sim 0.00013$.

\vspace{.3cm}
\noindent
\makebox[8cm][l]{plasma density $n\sim (25-30)\; {\rm cm}^{-3}$}  average magnetic field $\langle B\rangle \sim 9  \, \rm nT$,
\\
\makebox[8cm][l]{Alfv\'en speed $c_A = c \, \Omega_i/\omega_{pi} \sim 36 \, {\rm km/s}$,}
acoustic velocity $c_S \sim 50 \, {\rm km/s}$, \\
\makebox[8cm][l]{ion gyrofrequency $\Omega_i\sim 0.8 \, 1/\rm s$,} electron gyrofrequency $\Omega_e \sim 1.46 \times 10^3\, 1\rm s$\\
\makebox[8cm][l]{ion plasma frequency $\omega_{pi}\sim 6.6 \times 10^3\, 1/ \rm s$,} electron plasma frequency $\omega_{pe}\sim 280  \times 10^3\, 1/ \rm s$\\
\makebox[8cm][l]{ion plasma length $d_i=c/\omega_{pi} \sim 46 \, \rm km$,} electron plasma length $d_e=c/\omega_{pe} \sim 1 \, \rm km$\\
\makebox[8cm][l]{ion Larmor radius $\rho_{Li}=v_{Ti_\bot}/\Omega_i \sim (40-110) \, \rm km$,} electron Larmor radius $\rho_{Le}=v_{Te_\bot}/\Omega_e \sim (1-2.5) \, \rm km$ \\
\makebox[8cm][l]{ion beta $\beta_{i_\bot} = 2 p_{i_\bot}/c^2 \epsilon_0 B^2 \sim 0.5-2.5$,} electron beta $\beta_{e_\bot} = 2 p_{e_\bot}/c^2 \epsilon_0 B^2 \sim 1-2$
\\
\makebox[8cm][l]{ion temperature anisotropy $A_i = T_{i_\bot}/T_{i_\Vert} \sim 1.6$
,} electron temperature anisotropy $A_e = T_{e_\bot}/T_{e_\Vert} \sim 0.9$
\\
\makebox[8cm][l]{ion thermal velocity $v_{Ti}\sim 40 \, {\rm km/s}$,} electron thermal velocity $v_{Te} \sim 1500 \, {\rm km/s}$
%
%
%
\\
\makebox[8cm][l]{velocity of the slow solar wind $v_{sw}\sim 360 \, {\rm km/s} $,}
angle between $\langle\vec{B}\rangle$ and the solar wind $\theta_{BV}\sim 55^{\rm o}-125^{\rm o}$
\\
characteristic diameter of the structure transverse to the magnetic field $L_\bot \sim (5-25) \, c/\omega_{pi} \sim (6-30) \,\rho_{Li}$, \\
velocity (in the plasma frame) of the structure perpendicular to the magnetic field $u_\bot = (0.5-4) \, c_A \pm (1-4) \, c_A$.
\vspace{.3cm}

Relevant plasma parameters in 
the fast solar wind ($|\Omega_e|/\omega_{p e} \sim 0.013$, $\Omega_i/\omega_{p i} \sim 0.00030$) are \citep{2017ApJ...849...49P}:
\vspace{.2cm}

\noindent
\makebox[8cm][l]{plasma density $n\sim 4\; {\rm cm}^{-3}$}  average magnetic field $\langle B\rangle \sim 8.3  \, \rm nT$,
\\
\makebox[8cm][l]{ion temperatures $T_{i_\bot} = 41 \, {\rm eV}$, $T_{i_\Vert} = 30 \, {\rm eV}$ } 
electron temperatures $T_{e_\bot} = 18 \, {\rm eV}$, $T_{e_\Vert} = 14 \, {\rm eV}$,
\\
\makebox[8cm][l]{parallel ion thermal velocity  $v_{Ti_\Vert} \sim 53\, {\rm km/s}$,}
parallel electron thermal velocity  $v_{T e_\Vert} \sim 1600\, {\rm km/s}$, \\
\makebox[8cm][l]{Alfv\'{e}n speed $c_A = c \Omega_i/\omega_{pi} \sim 100 \, {\rm km/s}\sim 1.9 v_{Ti_\Vert} $,}
ion beta $\beta_{i_\bot} = 0.95$ 
\\
\makebox[8cm][l]{ion plasma length $c/\omega_{pi} \sim 115 \, {\rm km}$,} electron plasma length $c/\omega_{pe} \sim 2.7 \, \rm km$
\\
\makebox[8cm][l]{ion Larmor radius $\rho_{Li}=v_{Ti_\bot}/\Omega_i \sim 110 \, \rm km$,} electron Larmor radius $\rho_{Le}=v_{Te_\bot}/\Omega_e \sim 1.5 \, \rm km$ \\
\makebox[8cm][l]{velocity of the fast solar wind $v_{sw}\sim 600 \, {\rm km/s} $,}
angle between $\langle\vec{B}\rangle$ and the solar wind $\theta_{BV}\sim 50^{\rm o}-90^{\rm o}$

\section{Plasma parameters in the magnetosheath}\label{magsh}

Parameters in the magnetosheath region downstream of quasi-perpendicular bow shock ($|\Omega_e|/\omega_{p e} \sim 0.056$, $\Omega_i/\omega_{p i} \sim$ $ 0.0013$), and the properties of observed magnetic structures \citep{2004JGRA..109.5207A,2006JGRA..11112208A,2008NPGeo..15...95A}:

\vspace{.3cm}\noindent
\makebox[8cm][l]{plasma density $n\sim 30\; {\rm cm}^{-3}$}  average magnetic field $\langle B\rangle \sim 90  \, \rm nT$, \\
\makebox[8cm][l]{ion temperatures $T_{i_\bot} = 360 \, {\rm eV}$, $T_{i_\Vert} = 170 \, {\rm eV}$ } 
electron temperatures $T_{e_\bot} = 95 \, {\rm eV}$, $T_{e_\Vert} = 85 \, {\rm eV}$, 
\\
\makebox[8cm][l]{parallel ion thermal velocity  $v_{Ti_\Vert} \sim 130\, {\rm km/s}$,}
parallel electron thermal velocity  $v_{T e_\Vert} \sim 4000\, {\rm km/s}$, \\
\makebox[8cm][l]{Alfv\'{e}n speed $c_A = c \Omega_i/\omega_{pi} \sim 390 \, {\rm km/s}\sim 3 v_{Ti_\Vert} $,} 
ion beta $\beta_{i_\bot} = 0.5$
\\
\makebox[8cm][l]{ion plasma length $c/\omega_{pi} \sim 45 \, {\rm km}$,} electron plasma length $c/\omega_{pe} \sim 1 \, \rm km$\\
\makebox[8cm][l]{ion Larmor radius $\rho_{L i} = v_{Ti_\bot}/\Omega_i \sim 20 \, {\rm km}$,} electron Larmor radius $\rho_{Le} = (m_e T_{e\bot}/m_i T_{i\bot})^\frac{1}{2} \, \rho_{Li} \sim 0.25 \, \rm km$, \\
radius of the structure transverse to magnetic field $R_\bot\sim (400-500) \, {\rm km} \sim 10 \, c/\omega_{pi} \sim 20 \,\rho_{Li}$, \\
size of the structure along the magnetic field $L_\Vert > 1000 \, {\rm km}$, \\
velocities of the structure $\bot$ and $\Vert$ to the magnetic field $u_\bot = (35-100) \, {\rm km/s}$, \; $u_\Vert \sim (70-200) \, {\rm km/s} \sim v_{Ti_\Vert}$\\
bulk velocity of the plasma $v_{p_0}\sim 250 \, {\rm km/s} $.

\section{Equation of state}\label{eqstate}

{Here we demonstrate that in the regime studied in this paper the process may be approximately regarded as polytropic. Our generic equation of state (\ref{pressure}) has the  form:
\begin{equation}\label{pressureapp}
d p_\bot/p_\bot = \gamma_\bot \; dn/n , \quad\quad d p_\Vert/p_\Vert = \gamma_\Vert \; dn/n ,
\end{equation}
in which the multipliers $\gamma_\Vert$ and $\gamma_\bot$ are some functionals of the plasma density $n$. We consider only regimes in which the perturbations of the density and of the magnetic field are sufficiently small, see Eq. (\ref{weak_z_dep}) and the functionals $\gamma_\Vert$ and $\gamma_\bot$ can be estimated from the linearized Vlasov equation, viz.
\begin{equation}\label{vlasovapp}
\left[\frac{\partial}{\partial t} + \vec{v}\cdot\nabla + \frac{q}{m} \left(\vec{v}\times\vec{e}_z B_0\right)\cdot
 \frac{\partial}{\partial\vec{v}} \right]\delta f =
- \frac{q}{m}
\left(\vec{E}+\vec{v}\times\vec{B}\right)
\cdot\frac{\partial f_0}{\partial\vec{v}},
\quad{\rm where}\quad
f_0 = \frac{n_0}{(2\pi)^\frac{3}{2} v_{T_\Vert} v_{T_\bot}^2}\;\exp\left(-\frac{v_x^2+v_y^2}{v_{T_\bot}^2}-\frac{v_\Vert^2}{v_{T_\Vert}^2}\right)
\end{equation}
The unperturbed distribution $f_0$ is Maxwellian with anisotropic temperature, $T_\bot\ne T_\Vert$.
Using cylindrical coordinates in velocity space, $v_\bot=\sqrt{v_x^2+v_y^2}$, \; $\theta=\tan^{-1}(v_y/v_x)$, the linearized Vlasov equation takes the form
\begin{equation}\label{vlasovcylapp}
\left(\frac{\partial}{\partial t} + \vec{v}\cdot\nabla -\Omega\;\frac{\partial}{\partial\theta}\right)\delta f = \frac{q}{m}
\left(\nabla\phi - \nabla\times\nabla^{-2}\vec{B}-\vec{v}\times\vec{B}\right)
\cdot\frac{\partial f_0}{\partial\vec{v}},
\quad
{\rm where}\quad \Omega=\frac{q B_0}{m}.
\end{equation}
 Next, we apply the Fourier transformation in time, $\partial/\partial t\to -i\omega$, and in space, $\nabla\to i\vec{k}$, with $\vec{k}=k_y\vec{e}_y+k_z\vec{e}_z$, yielding
\begin{equation}\label{vlasovfourapp}
-i\left(\omega -k_z v_z - i\Omega\frac{\partial}{\partial\theta}\right)\left[\delta f\; \exp\left(-i\;\frac{k_y v_\bot}{\Omega}\;\sin\theta\right)\right] =
\frac{q}{m}\; f_0\;\left[v_\bot\left(u_x\sin\theta + u_y\cos\theta\right) + u_z v_z \right]\; \exp\left(-i\;\frac{k_y v_\bot}{\Omega}\;\sin\theta\right),
\end{equation}
where
\begin{eqnarray}\label{uxapp}
&& u_x\left(\omega, k_y, k_z, v_z\right) = \frac{B_z}{k_y}\left[\frac{\omega}{v_{T_\bot}^2} + v_z k_z\left(\frac{1}{v_{T_\Vert}^2}-\frac{1}{v_{T_\bot}^2}\right)\right], \\
\label{uyapp} && u_y\left(\omega, k_y, k_z, v_z\right) = \left(i k_y\phi+\frac{\omega k_z B_x}{k_y^2+k_z^2}\right)\frac{1}{v_{T_\bot}^2}+
v_z B_x\left(\frac{1}{v_{T_\Vert}^2}-\frac{1}{v_{T_\bot}^2}\right), \\
\label{uzapp} && u_z\left(\omega, k_y, k_z\right) =  \left(i k_z\phi-\frac{\omega k_y B_x}{k_y^2+k_z^2}\right)\frac{1}{v_{T_\Vert}^2}.
\end{eqnarray}
Noting that above exponential function is the generating function of Bessel functions, $\exp(i\zeta\sin\theta)=\sum_{l=-\infty}^\infty J_l(\zeta)\;e^{il\theta}$ and expanding the left hand side of Eq. (\ref{vlasovcylapp}) into cylindrical harmonics, we obtain
\begin{equation}\label{deltafapp}
\delta f = i\,\frac{q}{m}\, f_0\,\sum_{s-\infty}^\infty J_s\left(\frac{k_y v_\bot}{\Omega}\right)\,e^{is\theta} \times
\sum_{l=-\infty}^\infty\frac{e^{-il\theta}}{\omega-k_z v_z -l\theta}
\left(i\Omega u_x\,\frac{d}{dk_y} + \frac{l\Omega u_y}{k_y} + u_z v_z\right)\;
J_l\left(\frac{k_y v_\bot}{\Omega}\right),
\end{equation}
where we used the recurence relations for Bessel functions. As the unperturbed distribution function $f_0$ does not depend on the angle $\theta$, the above $\delta f$ is easily integrated in $\theta$, yielding the selection rule $\int_0^{2\pi} d\theta \, e^{i(s-l)\theta} = 2 \pi \delta_{s,l}$.}

{We also note that the functions $u_x$, $u_y$, $u_z$ do not depend on $v_\bot$, which permits an easy integration of the distribution function in $v_\bot$. In the calculation of the density perturbation, we will encounter the following integral
\begin{equation}\label{integralapp}
{\cal J}_l\left(k_y, v_{T_\bot}\right) \equiv \int_0^\infty v_\bot dv_\bot\, \exp\left(-\frac{v_\bot^2}{2 v_{T_\bot}^2}\right) \, J_l^2\left(\frac{k_y v_\bot}{\Omega}\right) = v_{T_\bot}^2 \,
\exp\left(-\frac{k_y^2 v_{T_\bot}^2}{\Omega^2}\right) \, I_l\left(\frac{k_y^2 v_{T_\bot}^2}{\Omega^2}\right),
\end{equation}
and its partial derivative $\partial {\cal J}_l(k_y, v_{T_\bot})/\partial k_y$. Likewise, in the calculation of the perturbation of the perpendicular pressure $p_\bot$ we will encounter the partial derivative $\partial {\cal J}_l(k_y, v_{T_\bot})/\partial(v_{T_\bot}^{-2})$.}

We perform the calculations with the accuracy to first order in the small-but-finite Larmor radius. Then, in the infinite sum we keep only the terms $l=0$, and $l=\pm 1$ and we expand the Bessel functions to first order in ${k_y^2 v_{T_\bot}^2}/{\Omega^2}$. The integration in $v_z$ is performed by expanding the functions $(\omega-k_z v_z)^{-1}$ and $[(\omega-k_z v_z)^2-\Omega^2]^{-1}$ in two limits, $\omega\gg k_z v_z$ and $\omega\ll k_z v_z$ and keeping the terms up to $k_z^2 v_z^2/\omega^2$ and $\omega^2/k_z^2 v_z^2$. Calculations are further simplified under drift scaling (\ref{drift_scaling}) and for a weak dependence along magnetic field lines, (\ref{weak_z_dep}). Under above conditions and after some lengthy calculations, we find that within the adopted accuracy the parallel functional $\gamma_\Vert$ reduces to a constant, viz. $\gamma_\Vert = 3$ when the characteristic parallel velocity of propagation  $u_z$ is bigger than the parallel thermal velocity $v_{T_\Vert}$ and the process can be considered as adiabatic, and to $\gamma_\Vert = 1$ when $u_z\ll v_{T_\Vert}$, i.e. the process is isothermal. Likewise, the perpendicular functional $\gamma_\bot$ reduces to $\gamma_\bot = 1$  for arbitrary ratios $u_z/v_{T_\Vert}$ if the characteristic perpendicular size of the solution is much bigger than the Larmor radius. Conversely, for solutions whose transverse scale approaches the ion scales, $\gamma_\bot$ can be approximated by a constant only in a limited number of cases, for which vortex solutions are found in Section \ref{Traveling}. These are the large-$\beta$ shear Alfv\'{e}n solution with $u_z\gg v_{T i_\Vert}$ whose parallel electric field is zero, $E_\Vert = 0$,
and the large-$\beta$ kinetic slow mode solution in the regime $c_A\gg u_z\gg v_{T i_\Vert}$, when the coupling with acoustic perturbation and the torsion of the magnetic field are negligible. In both cases we have $\gamma_{i_\bot} = 2 $. 

\begin{acknowledgements} This work was supported in part (D.J. and M.B.) by the MPNTR 171006 
and NPRP 11S-1126-170033 grants. D.J. and O.A. acknowledge financial support from the CIAS, from The French National Centre for Space Studies (CNES), and of the CNRS, and the hospitality of the LESIA laboratory in Meudon.
\end{acknowledgements}

\bibliography{Meudon_2014}

\end{document}